\documentclass[12pt]{JHEP3}
\usepackage{epsfig}
\usepackage{amsmath}
\newcommand{\p}{\partial}
\newcommand{\nn}{\nonumber}

\newcommand{\identity}{{\rlap{1} \hskip 1.6pt \hbox{1}}}
\title{The shape of nonabelian D-branes} 
  
\author{
Koji Hashimoto\\ 
Institute of Physics, University of Tokyo\\
Komaba, Tokyo 153-8902, Japan\\
E-mail: \email{koji@hep1.c.u-tokyo.ac.jp}
}

\abstract{
We evaluate bulk distribution of energies, pressures and various
D-brane/F-string charges generated by nontrivial matrix configurations
in nonabelian D-brane effective field theories, using supergravity
source density formulas derived originally in Matrix theory. 
Off-diagonal elements of worldvolume nonabelian fields, especially 
transverse scalar fields, induce various interesting bulk structures
exhibiting the shape of branes. First, we study the energy distribution
of string-brane networks generated in the bulk by the Yang-Mills
monopoles and the 1/4 BPS dyons, and confirm force balance of them. An
application to the Yang-Mills description of recombination of
intersecting D-branes gives results indicating presence of the tachyon
matter. Second, we analyse the shape of fuzzy D-branes given by
nonabelian scalar fields which are mutually noncommutative. We employ
fuzzy $S^2$, fuzzy $S^4$ and fuzzy cylinder/supertube as matrix
configurations of $N$ D0-branes representing higher dimensional
noncommutative D-branes. We find that in the continuum (large $N$) limit
the D-brane charge distributions become in the expected shape of a
sphere or a cylinder with an infinitesimal thickness. However, the
distributions found for finite $N$ are difficult to interpret, which
leaves a puzzle for a possible dual description in terms of  higher
dimensional D-branes. A resolution is provided with use of an ordering
ambiguity in the charge density formulas.
}
 

\preprint{
{\normalsize{\tt hep-th/0401043}}\\ 
{\normalsize UT-Komaba/04-01}
}

\begin{document}

\section{Introduction and summary}
\label{sec1}

The idea of {\it brane democracy}, originally used by P.~K.~Townsend
\cite{townsend}  
in the context of superalgebras of low energy supergravity in
string theories,  
has an interesting possibility to be a guiding principle to
seek for a unified description of string theories. Looking back the
developments in the last decade, as candidates for the unified
description, 
one may find Matrix theory and IIB matrix
model which adopt lower dimensional D-branes as fundamental
constituents, as well as tachyon condensation scenario of K-theory in
which one might start with the highest dimensional unstable
D-branes on the other hand. 
It is unclear why these descriptions based on D-branes with different
dimensions can be consistent, but we might have
relied substantially on the general idea of the brane democracy ---
all the physics in string/M-theory can be constructed by starting with
any dimensional D-branes.
The interplay between the above different descriptions
is still rather obscure,
leaving the question that what kind of objects in string theories 
can be consistently described in different scenarios.

One of the simplest examples in which a single brane object can be
described in two equivalent ways is a flat D2-D0 bound state. It can be
described by (i) a flat D2-brane with a constant magnetic field on it,
and also by (ii) infinite number of D0-branes whose representation
matrices satisfy Heisenberg algebra. These two descriptions give the
same boundary state \cite{ishibashi} and thus describe completely
identical physics. The essential part of the equivalence is that,
although the 
second description uses the D0-branes as constituents, their
matrix-valued 
configurations provide ``fuzzy'' plane  whose support is just a
flat $R^2$ which is identical with the geometry of the first
description.  

Next natural question which one might ask is to what extent this
``democratic'' duality on the flat D2-D0 bound state can be generalized. 
To answer this challenging question partly, 
in this paper we investigate geometric shapes produced by 
various fuzzy ($\sim$ matrix-valued, or noncommutative) configurations
of D0-branes (corresponding to (ii) in the above example), to see if
they are 
consistent with the expected dual descriptions in terms of higher
dimensional D-branes ((i) in the above).  
To extract the geometry of the fuzzy objects, we
look at low energy sector of closed string excitations whose sources
are generated by the matrix configurations of D0-branes. 
We make full use of the supergravity coupling
to the low energy D0-brane effective field theory derived in 
\cite{WT-Mark}. 

Even for a finite number of D0-branes, one can argue the existence
of this ``democratic'' duality in the following way. Let us consider 
$N$ D0-branes placed in a background constant RR (Ramond-Ramond) 4-form
field strength. As shown in \cite{myers}, these form a fuzzy $S^2$ which
is a classical solution of the D0-brane effective field theory in this
background. On the other hand, one finds a spherical D2-brane on which 
$N$ units of a magnetic field are turned on, as a classical solution of
a D2-brane effective field theory in this background. This has the
same number of the net D0-brane charge as the above fuzzy $S^2$
solution. Hence one can expect that these two might be the same object.  

The aim of this paper is to make full use of 
the supergravity charge density formulas to study the geometrical
information hidden in nonabelian configurations of D-brane worldvolume 
theories, and furthermore to see if this sort of expectation 
for the ``democratic'' duality with finite $N$ is realized or
not. We shall give a summary of the results of this paper in the
following. 

In section \ref{sec2} we review the supergravity charge density formulas
especially for D0-brane actions \cite{WT-Mark} and see how the
noncommutative (fuzzy) plane can give the consistent democratic duality
in them. 
The essence of the use of the supergravity charge density formulas
\cite{WT-Mark} is that one can extract physical and gauge-invariant
quantities from Matrix-valued, or equivalently, non-commutative or
fuzzy, configurations of nonabelian D-brane worldvolume theories.
Previously, the formulas were utilized to show the nontrivial brane
structures such as the presence of strings extended away
from D-brane worldvolumes, when nonabelian Yang-Mills configurations
including BPS monopoles are employed \cite{TH}. 
In section \ref{secene}, 
in order to show the power of the supergravity charge density formulas,
we compute bulk distribution of energy-momentum tensor for the
worldvolume solutions considered in \cite{HN} and \cite{TH}.
We start with a computation of the energy and the pressure
generated in the bulk by the solutions provided in \cite{HN} 
(a Yang-Mills description of recombination of intersecting D-branes).
This explicitly shows that, after the
recombination, there remains energy between the recombined D-branes
which was indicated in \cite{sato}. 
The pressure of this unknown ``matter'' tends to vanish as time
develops, hence we may regard this as an appearance of a 
tachyon matter \cite{tacmat} in Yang-Mills theory.
We computed also the energy and pressure distributions of the
supersymmetric solutions considered in \cite{TH}. It is shown that 
the presence of the D/F-strings away from the D-brane worldvolumes
is necessary for the force balance of the whole configurations.
For every part of the distributed strings, their charges and energies 
are consistent with the BPS bounds in the bulk with expected
supersymmetries.  
Furthermore, we calculate string charges in the bulk generated 
in 1/4 BPS dyons in 1+3 dimensional Yang-Mills theory \cite{su3,su4,
1/4d}. The curved shape of the D3-branes observed in \cite{su3,su4} 
is found to have an interpretation consistent with bulk $(p,q)$-strings
present away from the D3-brane worldvolume. 

A conclusion in section \ref{secene}
is that, with a single transverse scalar field (or mutually
commutative scalar fields), the supergravity charge
density formulas give consistent brane configurations in the bulk.
In the subsequent sections, we come to the main point of this paper :
the study of the shape of fuzzy D-branes defined by 
mutually noncommutative scalar fields.
We apply the RR density formulas to a fuzzy $S^2$, a fuzzy $S^4$, and a
fuzzy cylinder, in section \ref{sec3}, \ref{sec4} and \ref{sec5}
respectively. In section \ref{sec5}, we study also a fuzzy supertube 
as an application. We see that in the continuum limit ($\sim$ a large
$N$ limit) their RR charge distributions become in the shape of a sphere
(or a cylinder) with an infinitesimal thickness, as expected. However,
we find that for finite $N$, the charge distributions are different
from what we expect. For the finite $N$, the charges are on a collection
of shells with various radii, not on a single shell in the expected
shape. Interestingly, if we take the large $N$ (continuum) limit, these
shells annihilate with each other and only a single outer shell remains,
to give results consistent with the democratic duality.

The last section is devoted to discussions for 
possible interpretation of these strange distributions for the finite
$N$, as well as on the existence of the democratic duality at finite
$N$.  Assuming that the symmetrized ordering for the charge density
formulas may receive corrections of the form of commutators, we compute
the first nontrivial corrections by demanding the democratic duality. 

In the course of the study, it turns out that 
the supergravity charge density formulas for matrix configurations of
nonabelian worldvolume theories of D-branes are surprisingly powerful to
reveal the actual information of the brane configuration
in the bulk. We see various examples in which the formulas give new and
consistent brane configurations extended in the bulk space. 
For cases with a single scalar field (or with several scalars which are
commuting with each other), the formulas
give completely consistent pictures of bulk D-brane/F-string
configurations. 
However, for the cases with two or more noncommutative scalar fields 
the symmetrized trace prescription for the formulas 
doesn't seem to give bulk distribution consistent with the
democratic duality, and we may need some correction to the ordering in
the definition of the formulas.
Unfortunately on our first motivation for understanding of the precise 
democratic duality, the supergravity charge density formulas for the 
latter cases turn out to be not that helpful, 
but we find that the formulas
capture various intrinsic features of nonabelian
worldvolume physics of D-branes. A more precise formulation of the
correspondence between nonabelian worldvolume fields and the bulk
charge distribution will give us a hint 
for the democratic duality and more unified view of nonperturbative
string theory. 

%%%%%%%%%%%%%%%%%%%%%%%%%%%%%%%%%%%%%%%%%%%%%%%%%%%%%%%%%%%%%%
%%%%%%%%%%%%%%%%%%%%%%%%%%%%%%%%%%%%%%%%%%%%%%%%%%%%%%%%%%%%%%
%%%%%%%%%%%%%%%%%%%%%%%%%%%%%%%%%%%%%%%%%%%%%%%%%%%%%%%%%%%%%%

\section{Charge density formulas and fuzzy plane}
\label{sec2}

\subsection{Formulas}
\label{subsec2-1}

Any D-brane couples to supergravity fields as a source, since D-branes
are defined as boundaries of worldsheets and thus are sources of closed
strings. With a given D-brane configurations provided by vacuum
expectation values of the massless excitations of 
open strings ending on the D-brane, 
in principle one can compute how the D-brane is coupled to the closed
string excitations such as gravitons and RR gauge fields.
 
Here we review the supergravity 
charge density formulas derived in Matrix
theory \cite{WT-Mark} 
(see also \cite{kabat-WT,myers,schi,oo}). 
Let us consider a matrix configuration $X_i$ ($i=1,2,\cdots,9$)
which is 
the vacuum expectation value of the scalar fields coming from the 
excitation of the string connecting the D0-branes.
Then the D0-brane charge density in the momentum
representation is given by 
\begin{eqnarray}
\widetilde{J}_0(\vec{k})
= {\cal T}_{\rm D0}\;
{\rm Str} \left[
e^{ik_iX_i}
\right]
= {\cal T}_{\rm D0}\;
{\rm tr} \left[
e^{ik_iX_i}
\right].
\label{dens0}
\end{eqnarray}
where $k_i$ is a momentum in the target spacetime, and 
${\cal T}_{\rm D0}$ is the tension of the D0-brane.
Though the original definition uses symmetrized trace in which 
$X_i$ is treated as a unit of symmetrization, the exponential is already
in a symmetrized form, so we may use the normal trace to evaluate the
density. After an inverse Fourier transformation we will obtain the
actual D0-brane distribution in the bulk spacetime.

This formula can be interpreted also in terms of D-brane boundary
states. Consider an overlap of a boundary state of a D0-brane located at 
$x_i=0$ and a closed string state of $C_\mu^{\rm RR}$ (a RR 1-form gauge
field),  $\langle C^{\rm RR}_0(x_i) | B_{\rm D0}\rangle$.
This should be proportional to $\delta(x_i)$ (whose momentum
representation is a constant), since the boundary state 
is a source for closed string excitations. A generalization of the
boundary state for distributed D0-branes should be written as 
\begin{eqnarray}
 {\rm tr} \exp\left[
i\oint d\sigma \hat{P}_i(\sigma) X_i
\right] |B_{\rm D0}\rangle \ ,
\label{bs}
\end{eqnarray}
where $X_i$ are the matrices defining the bulk distribution 
of the D0-branes, and
$\hat{P}_i(\sigma)$ is a worldsheet current of spacetime translation
evaluated on a worldsheet boundary. 
In evaluating the overlap with the RR state, 
only the zero mode $k_i$ of $\hat{P}$ is effective, so
essentially the overlap is proportional to (\ref{dens0}) in the momentum
representation. 

We adopt the ordering of the matrices as specified by the
symmetrized trace \cite{kabat-WT} 
in the formula (\ref{dens0}) and in the following
formulas, but there is no a priori reason why  we
have to choose this symmetrized ordering \cite{Watireview}. 
However, as seen in the paper
\cite{TH} and as shown in the next section, for certain examples
with a single transverse scalar field this ordering prescription seems
to be completely consistent with the physical interpretation
of the resultant brane charge distributions. On the other
hand, when there are two or more nontrivial transverse scalars which are
noncommutative with each other, with the above ordering prescription we
encounter a strange result, as we will 
see in section \ref{sec3}, \ref{sec4} and \ref{sec5}. Until section 
\ref{sec6} where we consider a correction to the ordering, we shall use
the symmetrized trace. The symmetrized ordering
was confirmed also in some cases of 
string scattering amplitudes \cite{oo}. It is interesting to note that
in the pp-wave/Super Yang-Mills correspondence \cite{bmn} 
exactly the same kind of formulas appear with the symmetrized 
trace (see \cite{antal} for this aspect of the symmetrized
trace). This resemblance is not accidental but has an origin in 
the derivation of BMN operators from disk amplitudes \cite{imamura}.

The formula for D2-brane charge density generated by the 
configuration $X^i$ is
\begin{eqnarray}
 \widetilde{J}_{0ij} (\vec{k})
= \frac{{\cal T}_{\rm D0}}{2\pi\alpha'} 
{\rm Str} \left[
-i [X_i, X_j]e^{ik_lX_l}
\right] \ .
\label{dens2}
\end{eqnarray}
Here the symmetrization involves the commutator as another unit. 
It is easy to see that this symmetrized trace can be replaced by
the normal trace, again.
The D4-brane charge density is a bit more involved, 
\begin{eqnarray}
 \widetilde{J}_{0ijkl}(\vec{k})\equiv
\frac{{\cal T}_{\rm D0}}{2(2\pi\alpha')^2}
\epsilon_{ijklm}
\epsilon_{mpqrs}
{\rm Str}
\left(X_p X_q X_r X_s e^{ik_lX_l}\right).
\label{dens4}
\end{eqnarray}
In the parenthesis, $[X_i,X_j]$ should be regarded as a unit
for the symmetrization, through (a trivial generalization of) a
formula  
\begin{eqnarray}
 \epsilon_{pqrs5}X_pX_q X_r X_s
=
\left\{ [X_1,X_2], [X_3,X_4]\right\}
+
\left\{ [X_2,X_3], [X_1,X_4]\right\}
+
\left\{ [X_3,X_1], [X_2,X_4]\right\} \ ,
\nn
\end{eqnarray}
where $p,q,r,s=1,2,3,4$.

The coupling to the graviton gives the energy-momentum tensor
generated in the bulk. The formulas are
\cite{WT-Mark} 
\begin{eqnarray}
&& \widetilde{T}_{00}(\vec{k})  = {\cal T}_{\rm D0}
{\rm Str} \left[
\left(\identity + \frac12 \dot{X_i}^2 -\frac14 \frac1{(2\pi\alpha')^2}
[X_i,X_j]^2\right) e^{ik_lX_l}
\right] \, ,\\
&& \widetilde{T}_{0i}(\vec{k})  = {\cal T}_{\rm D0}
{\rm Str} \left[
\left(\!\dot{X}_i
\left(\!\identity\! +\! 
\frac12 \dot{X_i}^2 \!-\!\frac1{4(2\pi\alpha')^2}
[X_i,X_j]^2\!\right)
\right.\right.
\nn\\
&&\left.\left.
\hspace{70mm}
-\frac{1}{(2\pi\alpha')^2}[X_i,X_j][X_j,X_k]\dot{X}_k
\!\right) e^{ik_lX_l}
\right] \, , \nn\\
&& \widetilde{T}_{ij}(\vec{k})  = {\cal T}_{\rm D0}
{\rm Str} \left[
\left(\dot{X}_i\dot{X}_j - \frac1{(2\pi\alpha')^2}
[X_i,X_k][X_k,X_j]\right) e^{ik_lX_l}
\right] \, .
\end{eqnarray}
There should be higher order correction terms in this
expression, especially a term like $[X,X]^4$ in $T_{00}$, since naively
the zero momentum part ($k=0$) of the bulk energy-momentum tensor
$T_{00}$
may coincide with the Hamiltonian computed from 
a D0-brane effective field theory, that is, a nonabelian Born-Infeld
action plus its derivative corrections. 
The coupling to the NSNS (Neveu-Schwarz Neveu-Schwarz) $B$-field
provides the bulk distribution of the F-string charge, 
\begin{eqnarray}
 \widetilde{I}_{0i}(\vec{k}) 
&=& \frac{{\cal T}_{\rm D0}}{2\pi\alpha'}
{\rm Str} \left[
i[X_i,X_j]\dot{X}_j e^{ik_lX_l}
\right] \ . 
\label{Fch}
\end{eqnarray}
In these formulas the unit for the symmetrization is $[X_i,X_j]$,
$\dot{X}_j$ and $k_lX_l$. 

\subsection{Fuzzy plane}
\label{ncp}

Let us illustrate how the RR charge density 
formulas are treated in the simplest 
example of a fuzzy plane (noncommutative plane). The fuzzy plane is
defined by a nontrivial commutator 
\begin{eqnarray}
 [X_1, X_2]=i\theta.
\label{ncplane}
\end{eqnarray}
The representation of this algebra is infinite dimensional.
Let us see how this algebra generates a higher dimensional
charge. 
With use of an identity
\begin{eqnarray}
 {\rm tr} \left[
e^{ik\cdot X}
\right]
=\frac1{2\pi\theta} \delta^2(k) \ ,
\end{eqnarray}
The D2-brane charge density formula (\ref{dens2}) is
evaluated as
\begin{eqnarray}
 J_{012}(x) = {\cal T}_{\rm D2}
\displaystyle\Pi_{i=3}^9 \delta(x_i) \ .
\label{ncp2}
\end{eqnarray}
Here we used a relation among the tensions, 
${\cal T}_{\rm D2}={\cal T}_{\rm D0}/(4\pi^2 \alpha')$.
This (\ref{ncp2}), and the fact that other components of the D2-brane
charge vanishes,  explicitly shows that a flat D2-brane along $x_1$ and
$x_2$ directions is generated by the algebra (\ref{ncplane}). 
This result was used for the analysis of Seiberg-Witten map \cite{SW}. 

In addition, 
we can compute the D0-brane charge density (\ref{dens0}) as 
\begin{eqnarray}
 J_0(x) = \frac{{\cal T}_{\rm D0}}{2\pi\theta} 
\displaystyle\Pi_{i=3}^9 \delta(x_i) \ .
\label{ncp1}
\end{eqnarray} 
This means that smeared D0-branes bound on the generated D2-brane are 
present, and their density is given by $1/2\pi\theta$. From the (dual)
ordinary picture of the  D2-brane action, the induced D0-brane charge
density can be seen in the RR coupling as 
${\cal T}_{\rm D2}2\pi\alpha'F_{12}$, 
where $F_{12}$ is a constant gauge field strength on the flat D2-brane.
This coincides with (\ref{ncp1}) through a famous relation \cite{SW}
$F_{12}= 1/\theta$. 

Here we have seen only the supergravity couplings, but 
in this case it was
shown \cite{ishibashi} that couplings to all the closed string
excitations are coincident in the two pictures. In appendix A, 
we present an interesting realization of this democratic duality via
non-BPS D1-branes.

%%%%%%%%%%%%%%%%%%%%%%%%%%%%%%%%%%%%%%%%%%%%%%%%%%%%%%%%%%%%%%
%%%%%%%%%%%%%%%%%%%%%%%%%%%%%%%%%%%%%%%%%%%%%%%%%%%%%%%%%%%%%%
%%%%%%%%%%%%%%%%%%%%%%%%%%%%%%%%%%%%%%%%%%%%%%%%%%%%%%%%%%%%%%

\section{Energetics of strings between branes}
\label{secene}

In this section, to illustrate the power of the multipole moments
of the supergravity couplings in Matrix theory and in general
nonabelian 
D-brane actions, we compute the energy-momentum tensor generated in the
bulk space by the solutions of 1+1 and 1+3 dimensional nonabelian gauge 
theories. In \cite{TH}, 
the bulk interpretations of the Yang-Mills solutions with
nontrivial 
transverse scalar fields were given through the RR and NSNS $B$-field
couplings generated in the bulk. The solutions we employ in this paper
are -- 
$SU(2)$ solutions with or without preserving supersymmetries in 1+1
dimensions representing string networks \cite{TH} 
or recombined D-strings
\cite{HN},  and 
abelian/nonabelian BPS monopoles \cite{TH} and 
1/4 BPS dyons \cite{su3,su4} in 1+3 dimensions.
After showing the solutions and their RR and NSNS $B$-field couplings
which were provided partly in \cite{TH}, 
we compute the multipole moments in the 
coupling to the graviton to obtain the 
bulk distribution of the energy-momentum tensor generated by the matrix
configurations.

\subsection{Tachyon matter in Yang-Mills and D-brane recombination}
\label{sectacmat}

The power of the multipole moments of the supergravity charge densities
in
Matrix theory and nonabelian D-brane actions becomes manifest when one
considers condensation of off-diagonal components of the Yang-Mills
fields. One of the examples in which the off-diagonal condensation 
has a significant geometric
meaning was found in \cite{HN} where intersecting D-strings are
recombined by the condensation.
In this case the modes are tachyonic. Let us briefly look back this
phenomena of the tachyon condensation described solely by the Yang-Mills
theory. This 1+1 dimensional Yang-Mills theory has an $SU(2)$ gauge
group for the two D-strings, with its matter content consisting of a
gauge field 
$A_\mu = A^a_\mu (\sigma_a/2)$ ($a=1,2,3$) and a scalar field\footnote{
This $Y$ is normalized as $Y=2\pi\alpha' \Phi$ where $\Phi$ is a usual
YM scalar field, so that $Y$ represents a target space distance. We turn
on only a single scalar field among 8, for the present purpose.} $Y$. 
The classical solution representing
the D-strings intersecting at an angle $\theta$ is
\begin{eqnarray}
 Y = \tan(\theta/2) x \sigma_3 \ , \quad A_\mu = 0 \ ,
\label{backg}
\end{eqnarray}
where $x\equiv x^1$ is a worldvolume spatial direction. 
Note that $\theta=0$ corresponds to 
a parallel BPS pair of D-strings while $\theta=\pi$ represents 
an anti-parallel pair of D-strings (parallel brane-antibrane).  
The fluctuation eigen modes obtained in the Yang-Mills theory\footnote{A
fluctuation analysis for a  
T-dual configuration was provided in \cite{Aki-Wati}.}  
include the following tachyon of the Gaussian profile in $x$, 
\begin{eqnarray}
 Y(x)  
= 2\pi\alpha' C(t)\exp 
\left[\frac{m^2x^2}{2}\right](\sigma_1/2), 
\quad
 A_x(x)
=  C(x^0)\exp 
\left[\frac{m^2x^2}{2}\right](\sigma_2/2) \ .
\label{gauss}
\end{eqnarray}
Note that $m^2$ is negative here and in the following. 
The fluctuation field $C(t)$ obeys a free tachyon propagation
equation $(\p_t^2+m^2)(t)=0$ where the tachyonic mass squared is 
$m^2 = -\frac{1}{\pi\alpha'}|\tan(\theta/2)|<0$, which 
coincides\footnote{
In fact, the fluctuation spectrum obtained in this way in 1+1
dimensional Yang-Mills theory with 8 transverse scalar fields precisely
coincides with the low-lying part of the worldsheet spectrum.
It was pointed out in \cite{HN} that the first massive mode with 
$m^2 = \theta/2\pi\alpha'$ does not appear in the fluctuation of $Y$
nor $A_1$. It looks that this contradicts with the worldsheet spectrum,
but this is not the case. The fluctuation analysis of the remaining 7
transverse scalar fields provide 7 complex eigen modes at this mass
level, which coincides with the worldsheet results in \cite{bdl}.
In the same manner, the number of the fluctuations at each mass level
can be shown to be identical with that of corresponding 
string excitations. 
We thank S.~Nagaoka for bringing us this point.
} with
a string worldsheet spectrum given in \cite{bdl} for a small $\theta$.
One of the solutions of the tachyon propagation equation is 
a ``half S-brane'' 
\begin{eqnarray}
 C(t) = C e^{\sqrt{-m^2}t} \ .
\label{halfS}
\end{eqnarray}
This is an analogue of the half S-brane background considered in
\cite{half}, and in our context it 
means a classical evolution (recombination)
starts at the past infinity when the D-strings are intersecting. 
The other solutions include a ``full S-brane'' \cite{fullS} 
$C(t)=C\sinh (\sqrt{-m^2}t)$ in which the D-strings touch once at
$t=0$, and also an analogue of the rolling tachyon 
$C(t)=C\cosh (\sqrt{-m^2}t)$ considered originally in 
\cite{originalroll},  
for which the D-strings never touch each other. 
One can also consider a quantum treatment of this
development of the condensation as in \cite{sato}. 

Let us study the bulk charges generated by this matrix configuration
with the tachyon condensation $C(t)$. 
The bulk charge density formulas we need can be obtained by making a
T-duality on the formulas given in section \ref{sec2}.
For the RR two-form field, the formulas we need are
\begin{eqnarray}
\widetilde{J}_{01} &= &{\cal T}_{\rm D1} \mbox{Str} 
\left[e^{ik_yY}\right] \ ,  \label{D1current}
\\
\widetilde{J}_{0y} &= & {\cal T}_{\rm D1} \mbox{Str} \left[
e^{ik_yY}D_1 Y\right] \ ,
\nonumber
\end{eqnarray}
where we give a Fourier transform in which 
$k_y$ is the momentum along the $y$ direction, and Str is the
symmetrized trace, where $Y, D_1 Y$ (and $F_{01}$ in the later examples)
are treated as units in the symmetrization. For the NSNS $B$-field, 
\begin{eqnarray}
 \widetilde{I}_{01}(x,k_y)
&= &{\cal T}_{\rm D1}(2\pi\alpha')
{\rm Str}\left[e^{ik_yY}F_{10}\right] \ ,
\\
 \widetilde{I}_{0y}(x,k_y)
&= &{\cal T}_{\rm D1}(2\pi\alpha')
{\rm Str}\left[e^{ik_yY}F_{10}D_1Y\right] \ . 
\label{currentform}
\end{eqnarray}
A straightforward calculation shows that \cite{TH} the NSNS $B$-field
coupling
vanishes in the present solution with the tachyon condensation, and 
\begin{eqnarray}
J_{01} (x, y) &= &{\cal T}_{\rm D1} 
\bigl[\delta(y-\lambda(x)) + \delta (y+\lambda(x))\bigr] \ ,
\nn \\
J_{0y} (x, y)&=  &{\cal T}_{\rm D1} 
\lambda'(x)
\bigl[\delta(y-\lambda(x)) - \delta (y+\lambda(x))\bigr] \ , 
\label{d1chargedis}
\end{eqnarray}
where $\lambda(x)\equiv
\pi\alpha' \sqrt{m^4x^2+C(t)^2 e^{m^2x^2}}$ is the
magnitude of the eigenvalues of $Y$. Therefore the intersecting
D-strings are recombined \cite{HN} and its separation is given by the
magnitude of the tachyon condensation, $2\pi\alpha'C(t)$.

\begin{figure}[tp]
\begin{center}
\begin{minipage}{13cm}
\begin{center}
\includegraphics[width=6cm]{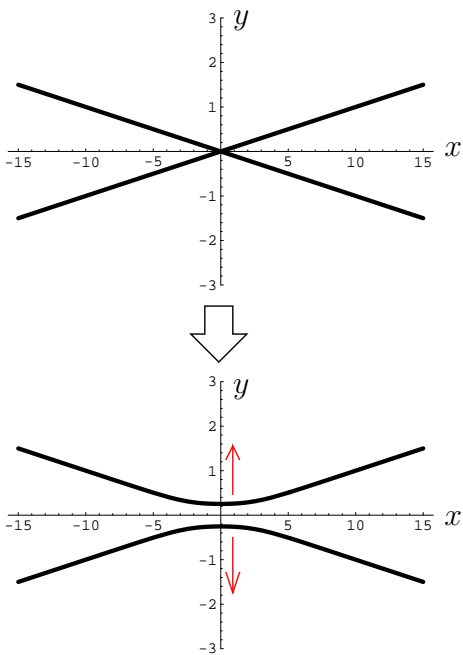}
\put(0,190){$x$}
\put(0,50){$x$}
\put(-80,240){$y$}
\put(-80,100){$y$}
\caption{The intersecting D-strings are recombined by the tachyon
 condensation. The small arrows indicate the motion of the D-strings.}
\label{tachyonfig}
\end{center}
\end{minipage}
\end{center}
\end{figure}

Let us turn our eyes to the graviton coupling which is given by
making the T-duality, 
\begin{eqnarray}
\widetilde{T}_{00} (x,k_y)
&= &{\cal T}_{\rm D1} \mbox{Str} 
\left[e^{ik_yY}\left(
\identity+ \frac12 (D_0Y)^2
+ \frac12 (D_1Y)^2
+ \frac12 (2\pi\alpha')^2 F_{01}^2
\right)\right] \ ,  
\\
\widetilde{T}_{11} (x,k_y)
&= &{\cal T}_{\rm D1} \mbox{Str} 
\left[e^{ik_yY}\left(
-\identity+ \frac12 (D_0Y)^2
+ \frac12 (D_1Y)^2
- \frac12 (2\pi\alpha')^2 F_{01}^2
\right)\right] \ ,  
\\
\widetilde{T}_{yy} (x,k_y)
&= &{\cal T}_{\rm D1} \mbox{Str} 
\left[e^{ik_yY}\left(
\frac12 (D_0Y)^2
- \frac12 (D_1Y)^2
\right)\right] \ .
\label{graviton}
\end{eqnarray}
In the expression we neglected higher order terms in fields
since we are working in Yang-Mills truncation. 
So we understand that the above formulas are reliable 
when these corrections are small. The order estimation \cite{HN}
shows that we have to work with 
\begin{eqnarray}
 C^2 \ll |m^2| \ .
\label{haveto}
\end{eqnarray}
With the 
tachyon condensation of the half S-brane (\ref{halfS}), this means that 
there is a limitation on $t$ beyond which the Yang-Mills analysis is not
reliable. 

In \cite{sato}, 
it was argued that in the recombination a part of
the energy in Yang-Mills may remain in the region between the recombined
D-strings. This is due to the non-vanishing energy 
originating in the off-diagonal gauge fields in the gauge
diagonalizing $Y$. The energy was interpreted 
to be due to creation of a bunch of F-string-anti-F-string pairs
connecting the upper and the lower D-strings.\footnote{The existence of
the strings connecting the recombined D-branes was observed in
\cite{tye}. } 
(Since the asymptotic F-string charge bound on the recombined D-strings
should vanish, the total number of created F-string charge should be
zero, which means that created string should appear as a pair of
F-string-anti-F-string.) 
This is a quite interesting phenomena, 
and here we apply the supergravity charge density 
formulas and confirm explicitly 
the presence of the string-like energy between the recombined D-strings.
Substitution of our Yang-Mills 
configuration into the above formulas for the energy-momentum tensor
gives
\begin{eqnarray}
 T_{\mu\nu} = T_{\mu\nu}^{(1)}
\left(\delta(y-\lambda) + \delta(y+\lambda)\right)
+ T_{\mu\nu}^{(2)}\left(\theta(y+\lambda) - \theta(y-\lambda)\right)
\ ,
\end{eqnarray}
where $\theta(z)$ is a step function which has a support only for $z>0$
as $\theta(z\!>\!0)=1$. The coefficients are given by
\begin{eqnarray}
 T_{00}^{(1)} /{\cal T}_{\rm D1}
&=& 1 + \frac{(\pi\alpha')^4}{2\lambda^2}
\left(C^2\dot{C}^2e^{2m^2x^2}+m^4x^2(m^2+C^2e^{m^2x^2})^2\right)
\\
T_{11}^{(1)} /{\cal T}_{\rm D1}
&=& -1 + \frac{(\pi\alpha')^4}{2\lambda^2}
\left(C^2\dot{C}^2e^{2m^2x^2}+m^4x^2(m^2+C^2e^{m^2x^2})^2\right)
\end{eqnarray}
\begin{eqnarray}
 T_{yy}^{(1)} /{\cal T}_{\rm D1}
&=&  \frac{(\pi\alpha')^4}{\lambda^2}
\left(C^2\dot{C}^2e^{2m^2x^2}-m^4x^2(m^2+C^2e^{m^2x^2})^2\right)
\\
 T_{00}^{(2)} /{\cal T}_{\rm D1}
&=& \frac{(\pi\alpha')^4}{2\lambda^3}e^{m^2x^2}
\left(2m^4x^2\dot{C}^2+C^2\dot{C}^2e^{m^2x^2}
+ C^2(m^2+C^2e^{m^2x^2})^2\right)
\label{preciseym}
\\
 T_{11}^{(2)} /{\cal T}_{\rm D1}
&=& \frac{(\pi\alpha')^4}{2\lambda^3}C^2e^{m^2x^2}
\left(-\dot{C}^2e^{m^2x^2}
+ (m^2+C^2e^{m^2x^2})^2\right)
\\
 T_{yy}^{(2)} /{\cal T}_{\rm D1}
&=& \frac{(\pi\alpha')^4}{\lambda^3}e^{m^2x^2}
\left(m^4x^2\dot{C}^2-C^2 (m^2+C^2e^{m^2x^2})^2\right)
\end{eqnarray}
where the dot denotes a derivative with respect to time $t$.
The important fact is that we have found a nonzero value of
$T^{(2)}_{\mu\nu}$. In fact, $T^{(2)}_{00}$ is the energy found between
the recombined D-strings.\footnote{The result we find here for
$T_{00}^{(2)}$ coincides with the number density of F-strings obtained
in \cite{sato}.} 
We find two more facts which support that this energy is due to
the creation of F-string pairs connecting the upper and the lower
recombined D-strings: first, the energy $T_{00}^{(2)}$ 
is homogeneous along the $y$ direction, and second, the pressure along
$y$ appears to be mostly negative, $T_{yy}^{(2)}<0$, while there is a
positive pressure along $x^1$, $T_{11}^{(2)}>0$. The second fact may
depend on 
the initial condition of the tachyon condensation, however if we employ
a typical case of the half S-brane (\ref{halfS}) we can show that this
is the case in the valid region of the rolling tachyon.
These strongly support the identification that the created energy is due
to some string-like object oriented along $y$ and connecting the two
D-strings, with no net charge.

\begin{figure}[tp]
\begin{center}
\begin{minipage}{13cm}
\begin{center}
\includegraphics[width=12cm]{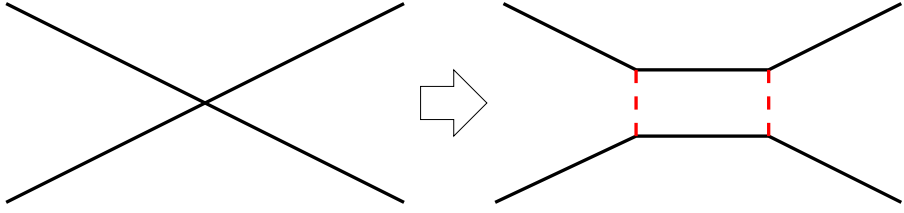}
%\put(0,190){$x$}
%\put(0,50){$x$}
%\put(-80,240){$y$}
%\put(-80,100){$y$}
\caption{A schematic view of the created strings (dashed lines)
 connecting the recombined D-strings (solid lines). In the actual
 Yang-Mills configuration the dashed lines are smeared horizontally,
 while here for the simplified calcuations presented in the text we
 draw localized vertical strings. }
\label{createfig}
\end{center}
\end{minipage}
\end{center}
\end{figure}

In fact, there is a reason why these strings should have appeared 
between the recombined D-strings. Imagine the situation in which 
these strings do not appear in the recombination. Then, the
intersecting D-branes can lower their energy by recombining and
separating from each other. The energy of them entirely comes from the 
lengths of the recombined D-strings. To evaluate this energy, 
we assume the shape of the recombined D-strings to be the solid lines in 
Fig.~\ref{createfig}. Defining the distance between the recombined
D-strings as $2\pi\alpha' C$, the energy gain due to the recombination
can be calculated as 
\begin{eqnarray}
\delta E = -T_{\rm D1}\pi\alpha'C \theta \ . 
\end{eqnarray}
Note that this is linear in $C$. This means that the instability of the
intersecting D-strings is the first order, that is, the pair of 
the intersecting D-branes is not a meta-stable vacuum in string theory.
If this were true, one cannot define the mass spectrum around this
configuration since this configuration is classically unstable.
However from the worldsheet analysis we know that the intersecting
D-branes are meta-stable vacua, hence we have to cancel this energy
$\delta E$ somehow. In fact, this energy linear in $C$ can be cancelled
by introducing a string-like ``bond'' between the recombined D-branes,
as in the dashed lines in Fig.~\ref{createfig}. The total tension of
this bond should be 
\begin{eqnarray}
 {\cal T} = |\delta E| / 2\pi\alpha' C = (\theta/2)T_{\rm D1} \ .
\end{eqnarray}
This is the total tension of the effective strings 
expected from this simple argument of
the meta-stable vacuum. Surprisingly, one can show that the precise
Yang-Mills result for $T_{00}^{(2)}$ (\ref{preciseym}) reproduces this
effective string 
tension, 
\begin{eqnarray}
 \lim_{C\rightarrow 0}\int\!\! dx\; 
T_{00}^{(2)}= (\theta/2)T_{\rm D1} \ .
\end{eqnarray}
To show this explicitly, we have used the half S-brane background
(\ref{halfS}). The integrand in the left hand side of this equation
is a delta function $\delta(x)$ in the limit $C\rightarrow 0$,
which means that the created strings are localized at the intersection
point before the tachyon condensation. These strings play the role of
the ``bond'' which prevents recombined D-strings from coming apart.

\begin{figure}[tp]
\begin{center}
\begin{minipage}{7cm}
\begin{center}
\includegraphics[width=6cm]{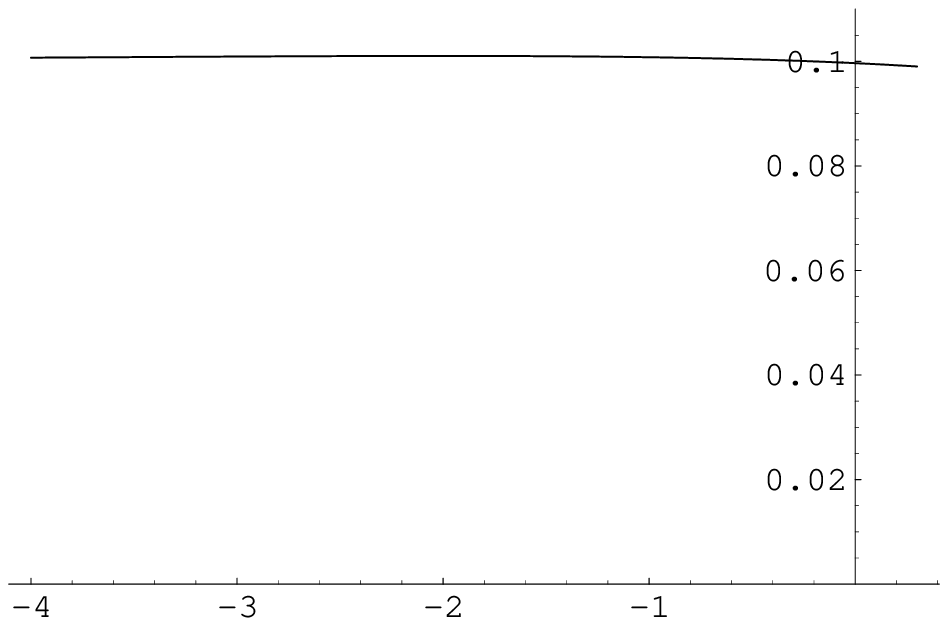}
\put(-10,70){$\displaystyle\int\!dx T_{00}^{(2)}$}
\put(0,10){$t$}
\caption{The energy resides between the recombined D-strings, integrated
 over $x^1$. We took $m^2=-0.1$ and a half S-brane 
 $C(t)=0.1 e^{\sqrt{-m^2}t}$ in the unit $\pi\alpha'=1$.}
\label{energyfig}
\end{center}
\end{minipage}
\hspace*{5mm}
\begin{minipage}{7cm}
\begin{center}
\includegraphics[width=6cm]{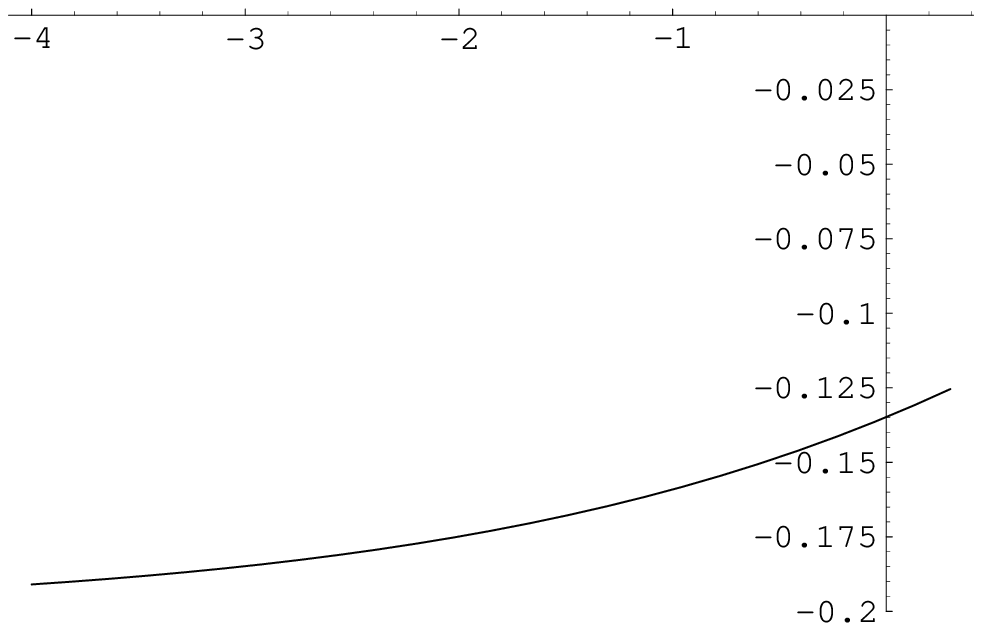}
\put(-10,60){$\displaystyle\int\!dx T_{yy}^{(2)}$}
\put(0,100){$t$}
\caption{The pressure $T_{yy}^{(2)}$ between the recombined D-strings,
 integrated over $x^1$, with the same parameters. Its magnitude tends to
 decrease.} 
\label{pressurefig}
\end{center}
\end{minipage}
\end{center}
\end{figure}

Once the pairs of F-strings are created between the recombined
D-strings, they may decay to something since they are unstable.
However we are working in Yang-Mills theory where the coupling to the
closed strings is neglected, so the energy would remain there. 
This leads us to the expectation that 
the end product should be a tachyon matter \cite{tacmat}, since
as studied in detail in \cite{HN}, the recombination of the intersecting
D-strings is a local version of the tachyon condensation of
a brane-antibrane. The tachyon matter was originally considered as an
end-product of a homogeneous tachyon condensation on a
brane-antibrane or a non-BPS brane \cite{tacmat}. 
It is quite intriguing that we can observe the tachyon matter even in 
the scheme of Yang-Mills theory. 
In fact, we can give an evidence for that the end product is the tachyon
matter. 
The important feature of the tachyon matter is that it is pressure-less
in the far future ($t \rightarrow \infty$) while it preserves a nonzero
energy density. In our case we 
cannot follow the large $t$ behaviour in the Yang-Mills theory, but we
may work in the region where (\ref{haveto}) is satisfied. The energy is
now distributed in the 
two-dimensional region between the D-strings, especially has a
dependence on the $x^1$ direction. In the 
correspondence to the brane-antibrane where $\theta=\pi$, this $x^1$
direction would shrink effectively \cite{HN} and thus to see the
quantity corresponding to the brane-antibrane picture we may
integrate over the $x^1$ direction. Interestingly, the energy
$T_{00}^{(2)}$ integrated over $x^1$ looks conserved, as in the case of
the rolling tachyon on the brane-antibrane, see
Fig.~\ref{energyfig}.\footnote{In this numerical evaluation we took 
the region $t<0$, since at this boundary $t=0$ one has 
$|C^2/m^2|=0.1$ and beyond
this point our Yang-Mills analysis may be violated.}
On the other hand, the magnitude of the 
pressure along the $y$ direction which is the
worldvolume direction of the brane-antibrane, $T_{yy}^{(2)}$, has a
tendency to decrease as time develops, after the integration over $x^1$
(see Fig.~\ref{pressurefig}). So the Yang-Mills result is consistent
with the brane-antibrane picture, and we conclude that Yang-Mills
analysis describing the recombination of the D-strings captures also the
property of the tachyon matter.

The energy $T_{00}^{(1)}$ is localized on the recombined D-strings, and 
it provides a time-dependent tension of them. Since we know the motion
of the D-strings which is given by $y=\lambda(t,x)$, let us compare
this $T_{00}^{(1)}$ with the energy of a single moving D-string whose
location is specified by $\lambda$. The hamiltonian of a single D-string
is given by
\begin{eqnarray}
 H_{\rm D1} = {\cal T}_{\rm D1}
\frac{1 + \lambda'^2}{\sqrt{1-\dot{\lambda}^2+\lambda'^2}}
= {\cal T}_{\rm D1}\left[
1 + \frac12 \lambda'^2 + \frac12 \dot{\lambda}^2 
+ \mbox{higher terms} \right] \ ,
\end{eqnarray}
where the dash denotes a derivative with respect to $x\equiv x^1$.
The ``higher terms'' consist of the Born-Infeld corrections and higher
derivative corrections in the abelian D1-brane action.
Interestingly, one can show that this hamiltonian 
coincides with the energy-momentum
tensor found above,
\begin{eqnarray}
 H_{\rm D1} = T_{00}^{(1)}.
\end{eqnarray}
This is a strong support for that the energy-momentum tensor formulas
(\ref{graviton}) provide a correct distribution of the 
energy in the bulk for the recombination of the D-strings.

\subsection{Energetics of a string network}

\begin{figure}[tbp]
\begin{center}
\begin{minipage}{13cm}
\begin{center}
\includegraphics[width=6cm]{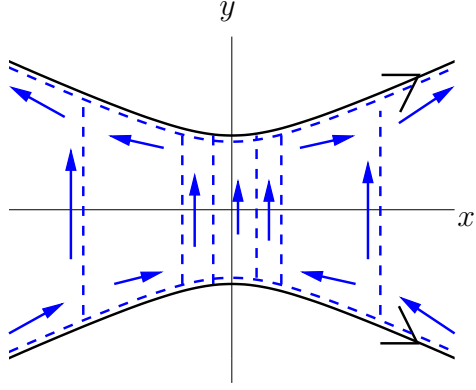}
\put(-90,140){$y$}
\put(0,60){$x$}   
\caption{Charge distributions in the bulk produced by the Yang-Mills
 solution representing a string network. Arrows indicate the directions
 of the flux for the fundamental strings. The dashed lines denote the
 charge itself. Note that the vertical dashed lines are actually smeared
 along $x$ although they look localized in this figure. Solid lines
 denote the $(p,q)$-strings.}   
\label{juncfig}
\end{center}
\end{minipage}
\end{center}
\end{figure}

A solution representing a supersymmetric recombination of
$(p,q)$-strings was found in \cite{TH} where, by calculating the
distributions of the D-string and fundamental string charges in the
bulk, it was found that the fundamental strings connect the recombined
D-strings. This is a generalization of the string network, and a naive
estimation of the force balance was given there. Here we explicitly
evaluate the energy-momentum tensor generated by this solution and show
the stability and the BPS bound. The supersymmetric 
solution in 1+1 dimensional Yang-Mills theory is quite simple \cite{TH}, 
\begin{eqnarray}
 Y=2\pi\alpha' A_0 = px\sigma_3 + a \sigma_1, \quad A_1 =0 \ ,
\label{simplesol}
\end{eqnarray}
where $p$ and $a$ are constant parameters which can be taken to be
non-negative without losing generality. The generated 
F-string charge is given as \cite{TH}
\begin{eqnarray}
 I_{01}(x,y)
&=&
{\cal T}_{\rm D1}\lambda'
\bigl[\delta(y-\lambda) - \delta (y+\lambda)\bigr] \ ,
\label{eq:p1}
\\
 I_{0y}(x,y)&= & {\cal T}_{\rm D1}\lambda'^2 \bigl[
\delta(y-\lambda) + \delta(y+\lambda)
\bigr]
+ {\cal T}_{\rm D1}\lambda'' \bigl[
\theta(y+\lambda) - \theta(y-\lambda)
\bigr] \ .\label{eq:2-current}
\end{eqnarray}
while the D-string charge has the same expression as (\ref{d1chargedis})
except that now the function specifying the location $\lambda(x)$ is
defined as $ \lambda(x) \equiv \sqrt{p^2 x^2 + a^2}$. The step function
term in (\ref{eq:2-current}) shows the presence of the F-string directed
along $y$ direction and connecting the recombined $(p,q)$-strings. 
The configuration is shown in Fig.~\ref{juncfig}. 

A simple calculation shows that 
the bulk energy-momentum tensor is generated as 
\begin{eqnarray}
 T_{00} &=&{\cal T}_{\rm D1}(1+\lambda'^2) 
\left(\delta(y-\lambda) + \delta(y+\lambda)\right)
+ {\cal T}_{\rm D1}\lambda''^2\left(\theta(y+\lambda) 
- \theta(y-\lambda)\right)
\ , \;
\\
 T_{11} &=& - {\cal T}_{\rm D1}
\left(\delta(y-\lambda) + \delta(y+\lambda)\right)
\ ,
\label{pre1}
\\
 T_{yy} &=&- {\cal T}_{\rm D1}\lambda'^2
\left(\delta(y-\lambda) + \delta(y+\lambda)\right)
- {\cal T}_{\rm D1}
\lambda''^2\left(\theta(y+\lambda) - \theta(y-\lambda)\right)
\ .
\label{pre2}
 \end{eqnarray}
The first term in $T_{00}$ describes the tension of the recombined
$(p,q)$-strings. Noting that the expression is provided with the delta
function of $y$, we take into account the slope $\lambda'$ and find that
the tension of it is ${\cal T}_{\rm D1}\sqrt{1+\lambda'^2}$. 
This in fact saturates the BPS bound, since the 
F-string density vector derived from (\ref{eq:p1}) 
and the first terms in (\ref{eq:2-current}) 
is ${\cal T}_{\rm D1}\lambda' \vec{n}$ where $\vec{n}$ is the unit
vector along the trajectory of the $(p,q)$-string 
(we took into account the
slope factor again). The second term in $T_{00}$ is the tension of the
F-string connecting the recombined $(p,q)$-strings, and again saturates
the BPS bound since it coincides with the density of the connecting 
F-strings provided in the second term in (\ref{eq:2-current}).
The orientation of all of these strings are found to be consistent with
the bulk supersymmetry of string networks studied in \cite{das}.

Let us check the force balance from the pressure terms
(\ref{pre1}) and (\ref{pre2}). First, the force along the $x$ direction
exists only on the recombined $(p,q)$-strings since there is no step
function term in (\ref{pre1}). This is consistent with the fact that the
connecting F-strings are oriented along $y$ direction. The net force
along the $x$ direction is given by 
\begin{eqnarray}
 \int_{x=\bar{x}} 
\!\!\!dy\; T_{11} = -2 {\cal T}_{\rm D1}
\end{eqnarray}
which is constant and does not depend on a slice 
$\bar{x}$. The support of the
integrand is a delta function whose location is just at the intersection
of the D-strings and the integration slice.
This means that the force
along the $x$ direction is balanced. The more nontrivial check is on the
force along $y$. We find 
\begin{eqnarray}
 \int_{y=\bar{y}} \!\!\! dx\; T_{yy} = -2p {\cal T}_{\rm D1}
\label{forcey}
\end{eqnarray}
and this is independent of $x$, which shows the force balance. Note that
the net force is carried only by the connecting F-strings if 
$|\bar{y}|<a$ while only by the $(p,q)$-strings if 
$|\bar{y}|\rightarrow \infty$. So the 
force is transfered from one to another when one changes the slice
$\bar{y}$, while the net force is kept fixed. This is the same as the
well-known stability of BPS string networks. 

The force balance condition is intimately related to the current
conservation of the energy-momentum tensor. For any solution of Matrix
theory, it was shown that 
the current conservation for bulk currents including the energy
momentum tensor holds \cite{currentcon}. Our system is almost 
T-dual to the Matrix theory (D0-brane effective theory), thus in our
case we may expect the current conservation. For example, we have
\begin{eqnarray}
 \p_y T_{yy} + \p_1 T_{1y} - \p_0 T_{0y}=0 \ .
\end{eqnarray}
Noting that the configuration is static ($\p_0 T_{\mu\nu}=0$), we
integrate this expression over $x$ and obtain
\begin{eqnarray}
 \p_y \int T_{yy} dx =0 \ .
\label{cucn}
\end{eqnarray}
We assumed that there is no surface term for $T_{1y}$. The result is
consistent with the equation (\ref{forcey}). However, 
we find that this
current conservation condition (\ref{cucn}) is not so strong to give the
force balance, since (\ref{cucn}) means just that $\int dx T_{yy}$ is
independent of $y$ and thus says that it may depend on $x$. 
The actual force balance is established by showing that the quantity such
as $\int dx T_{yy}$ is constant for each connected distribution of the
energy-momentum in the bulk.

\subsection{The shape of monopoles}

In \cite{TH}, the RR charge distributions of the BPS
monopole configurations in 1+3 dimensional 
$U(1)$ and $SU(2)$ Yang-Mills theory
were computed, and in the $SU(2)$ case the existence of D-strings
extending out of the D3-brane worldvolume has been found. 
Here we employ the same classical solutions and evaluate the
energy-momentum tensors generated in the bulk, to see the force balance
and the consistency with the RR densities found in \cite{TH}.

The formulas we need here are obtained by taking T-dualities third times
on the formulas for D0-branes. The result for the case
$D_0Y=F_{0i}=0$ is
\begin{eqnarray}
\widetilde{T}_{00} (x_i,k_y)
&= &{\cal T}_{\rm D3} \mbox{Str} 
\left[e^{ik_yY}\left(
\identity+ \frac12 (D_iY)^2
+ \frac14 (2\pi\alpha')^2 F_{ij}^2
\right)\right] \ ,  
\\
\widetilde{T}_{yy} (x_i,k_y)
&= &- {\cal T}_{\rm D3} \mbox{Str} 
\left[e^{ik_yY}
 (D_iY)^2
\right] \ ,  
\\
\widetilde{T}_{11} (x_i,k_y)
&= &{\cal T}_{\rm D3} \mbox{Str} 
\left[e^{ik_yY}\left(
-\identity+ \frac12 \left((D_1Y)^2-(D_2Y)^2-(D_3Y)^2\right)
\right.\right.
\nn\\
& & \hspace{40mm}\left.\left.
+ \frac12 (2\pi\alpha')^2 \left(F_{12}^2+F_{13}^2-F_{23}^2\right)
\right)\right] \ ,  
\end{eqnarray}
where $i=1,2,3$ is the worldvolume coordinates of the D3-brane(s).
Similar formulas can be obtained for $T_{22}$ and $T_{33}$.
In the examples we use in the following, the BPS equation 
\begin{eqnarray}
 2\pi\alpha' \epsilon_{ijk}F_{ijk} = 2D_i Y
\end{eqnarray}
holds, and the above formulas reduce to
\begin{eqnarray}
&&
\widetilde{T}_{00} (x_i,k_y)
= {\cal T}_{\rm D3} \mbox{Str} 
\left[e^{ik_yY}\left(
\identity+ (D_iY)^2
\right)\right] \ ,  
\nn\\
&&
\widetilde{T}_{yy} (x_i,k_y)
= - {\cal T}_{\rm D3} \mbox{Str} 
\left[e^{ik_yY}
 (D_iY)^2
\right] \ ,  
\nn\\
&&
\widetilde{T}_{11} (x_i,k_y)
=
\widetilde{T}_{22} (x_i,k_y)
=
\widetilde{T}_{33} (x_i,k_y)
= -{\cal T}_{\rm D3} \mbox{Str} 
\left[e^{ik_yY}\right] \ .  
\label{enemono}
\end{eqnarray}

\begin{figure}[tp]
\begin{center}
\begin{minipage}{10cm}
\begin{center}
\includegraphics[width=6cm]{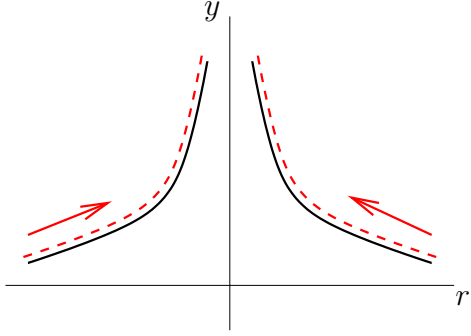}
\put(-95,120){$y$}
\put(0,10){$r$}
\caption{Charge distribution in the BIon. 
The dashed lines with arrows are the
 D-string charges bound on the D3-brane spike (solid lines).}
\label{bionfig}
\end{center}
\end{minipage}
\end{center}
\end{figure}

First we deal with the abelian monopole (BIon) \cite{CM}, 
\begin{eqnarray}
 B^i =\frac{1}{2} \epsilon_{ijk}F_{jk}
=\frac{-bx_i}{ r^3}, \quad Y = 2\pi\alpha'\frac{b}{r},
\quad r \equiv \sqrt{(x_1)^2+(x_2)^2 + (x_3)^2} \ .
\label{eq:abelian-monopole} 
\end{eqnarray}
The generated D-string density is \cite{TH}
\begin{eqnarray}
 J_{0y} &=& 
{\cal T}_{\rm D3}
(2\pi\alpha')^2
\frac{b^2}{r^4}\delta(y-2
\pi\alpha'
b/r) \ ,
\\
 J_{0i}&=& 
-{\cal T}_{\rm D3}
(2\pi\alpha')
 \frac{bx_i}{r^3}\delta(y- 2
\pi\alpha'
b/r) \ ,
\end{eqnarray}
which is depicted in Fig.~\ref{bionfig}. Since the theory is abelian,
the location of the 
D3-brane charge is trivially determined by the value of $Y$. The
computation of the energy-momentum tensor (\ref{enemono}) results in
\begin{eqnarray}
 T_{00} &=& {\cal T}_{\rm D3} \left(
1 + (2\pi\alpha')^2\frac{b^2}{r^4}
\right)\delta(y-2\pi\alpha' b/r) \ ,
\\
 T_{yy} &=& -{\cal T}_{\rm D3} 
 (2\pi\alpha')^2\frac{b^2}{r^4}
\delta(y-2\pi\alpha' b/r) \ ,
\\
 T_{11}= T_{22}= T_{33} 
&=& -{\cal T}_{\rm D3} \delta(y-2\pi\alpha' b/r) \ .
\end{eqnarray}

As described in the previous subsection, the force balance can be seen
in the integration of the energy-momentum tensor. That for the 
$x_i$ directions is trivial, 
\begin{eqnarray}
 \int_{x_1=\bar{x}_1} \!\!\!\!dydx_2dx_3\; T_{11} 
= -{\cal T}_{\rm D3} V_{23} 
\end{eqnarray}
where $V_{23}$ is the volume of the $x^2$ and $x^3$ direction. This is
independent of $x^1$ which shows the force balance along $x^1$. A 
physically important check is the force balance along $y$, which can be
shown as 
\begin{eqnarray}
 \int_{y=\bar{y}} \!\!\!\! dx_1dx_2dx_3\; T_{yy} 
= -4\pi {\cal T}_{\rm D3}2\pi\alpha' b = -n {\cal T}_{\rm D1}
\end{eqnarray}
which coincides with the tension of $n$ D-strings where the number of
the D-strings has been quantized as $n=b/2$. One can show that the
$T_{00}$ saturates a BPS bound of the D1-D3 bound state and also that
the orientation of the surface is consistent with the bulk
supersymmetry, as in the same way as the string network of the 
previous section.

\begin{figure}[tp]
\begin{center}
\begin{minipage}{10cm}
\begin{center}
\includegraphics[width=8cm]{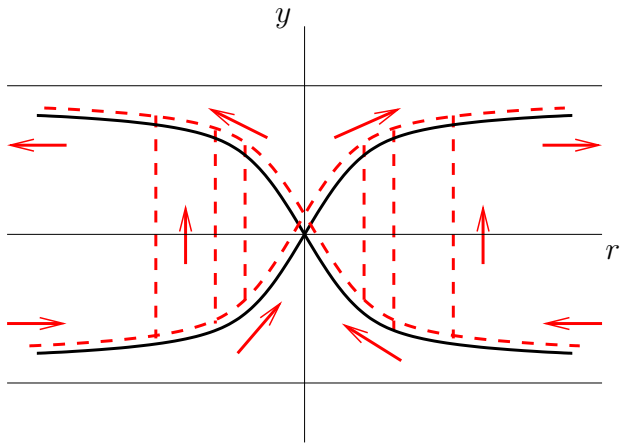}
\put(-125,160){$y$}
\put(0,70){$r$}
\caption{D-string charge distribution (dashed lines) of the 
'tHooft-Polyakov  monopole. D-strings away from the D3-brane surface
 connect the upper and the lower D3-branes.}
\label{f:monopole}
\end{center}
\end{minipage}
\end{center}
\end{figure}

The power of the multipole moment formulas becomes manifest when we
consider nonabelian monopoles. The $SU(2)$ 'tHooft-Polyakov BPS monopole
solution is \cite{Prasad-Sommerfield}
\begin{eqnarray*}
A_i  =   \epsilon_{aij} (1-K (r)) \frac{x_j}{r^2} (\sigma_a/2)
\ , \quad
A_0  =   0 \ , \quad 
Y= -2\pi\alpha' H (r) \frac{x_i}{r^2}(\sigma_i/2)  \ ,\nonumber
\end{eqnarray*}
where
$K (r)  \equiv Pr/\sinh (Pr)$, $H (r)  \equiv   Pr\coth (Pr) -1$,
and $P$ is a constant. In \cite{Aki}, 
this configuration was interpreted to be a
D-string suspended between parallel D3-branes separated by
$2\pi\alpha'P$, through the matching of the supersymmetry, 
the total energy and the RR charges. Here we can use the charge density 
formulas to
evaluate the actual bulk distribution of the energy and the
charges. Using (\ref{enemono}),  we obtain
\begin{eqnarray}
&& T_{00}/{\cal T}_{\rm D3}
= 
\left(1\! +\! (2\pi\alpha')^2\frac{(1\!-\!K^2)^2}{4r^4}\right)
\left(\delta(y\!-\!\lambda) + \delta(y\!+\!\lambda)\right)
\nn\\&& \hspace{60mm}
+ 2\pi\alpha'\frac{HK^2}{r^3}
\left(\theta(y\!+\!\lambda) - \theta(y\!-\!\lambda)\right) ,
\nn
\\
&& T_{11}/{\cal T}_{\rm D3}
=T_{22}/{\cal T}_{\rm D3}=T_{33}/{\cal T}_{\rm D3}
= 
-\left(\delta(y\!-\!\lambda) + \delta(y\!+\!\lambda)\right) \ ,
\label{enedism}
\\
&& T_{yy}/{\cal T}_{\rm D3}
= 
-(2\pi\alpha')^2\frac{(1\!-\!K^2)^2}{4r^4}
\left(\delta(y\!-\!\lambda) + \delta(y\!+\!\lambda)\right)
\nn\\&& \hspace{60mm}
- 2\pi\alpha'\frac{HK^2}{r^3}
\left(\theta(y\!+\!\lambda) - \theta(y\!-\!\lambda)\right) \ .
\nn
\end{eqnarray}
The D-string charge distribution was computed in \cite{TH}, 
\begin{eqnarray}
 J_{0y} = - T_{yy}, \quad 
 J_{0i}  =   (2\pi\alpha') {\cal T}_{\rm D3}\frac{x_i}{2r^3}
(1-K^2)
\left(\delta(y-\lambda) - \delta (y+\lambda)\right) \ .
\label{eq0a}
\end{eqnarray}
Here $\lambda$ denotes the location of the D3-brane which is specified
by the eigenvalues of $Y$, $\pm y =-\lambda(x_1,x_2,x_3) \equiv
\pi\alpha'H(r)/r$. 
The D-string charge distribution is depicted in Fig.~\ref{f:monopole}. 
As for the first term in $T_{00}$, one can show in a similar way that
this provides a BPS mass bound where the number of bound D-strings can
be found in the $J^{0i}$ above. The second term, which exists only
between the two D3-branes, is consistent with the D-string
distribution since it coincides with the second term in
$J^{0y}$. Therefore this configuration is supersymmetric in the bulk. 

Let us compute the force balance. Along the $y$ direction, we find 
\begin{eqnarray}
 \int_{y=\bar{y}}\!\!\! d^3x\; T_{yy} = -{\cal T}_{\rm D1}
\end{eqnarray}
for any slice with fixed $y$. Therefore, we conclude that 
the force is balanced. To show this the presence of the bulk D-strings
are necessary, thus the bulk D-strings are pulling the
D3-brane surfaces to make them in the shape determined by the
eigenvalues of the field $Y$. 

An additional observation with the
above equation is that the total pressure is equal to that of a single
D-string. Hence one can say that if we neglect the detailed structure
around the origin, macroscopically this configuration is equivalent to a
D-string suspended between two parallel D3-branes as interpreted in
\cite{Aki}. Let us study this in detail.\footnote{We thank A.~Hashimoto
for a discussion on this point.} With the slice $y=0$, all the
D-string charge in this configuration is provided by the bulk D-strings,
that is, the second terms in (\ref{enedism}). The distribution of the
bulk D-string is given by the number density, $2\pi\alpha'H K^2/r^3$.
This function has a single peak at $r=0$ and it damps for large $r$ as
$e^{-2Pr}$. This means that, when the asymptotic separation of the
D3-brane ($=2\pi\alpha' P$) is large compared to the string scale
($\Leftrightarrow P \gg 1/l_{\rm string}$), the bulk D-string
charge is distributed within a tube with a radius smaller than the
string scale. 
This is consistent with \cite{Aki} in that for the large separation the
detailed structure of the distributed D-strings in the bulk
is hidden in the region within the distance of the string length 
from the D-brane surfaces.
If we take $P\rightarrow \infty$ limit (the large separation between the 
asymptotically parallel D3-branes), the bulk D-strings $HK^2/r^3$
disappear, which is consistent with the the abelian case 
we considered before. 

\subsection{The shape of 1/4 BPS dyons}
\label{1/4}

In \cite{su3}, a new solution of ${\cal N}=4$
$SU(3)$ Yang Mills theory in $1+3$ dimensions was
constructed. The solution preserves $1/4$ of the original
supersymmetries and possesses a dyonic charge where the electric charge
is not parallel to the magnetic charge due to the inclusion of two
transverse scalar fields $Y$ and $Z$. The existence of this kind of
solutions was conjectured in \cite{bergman} 
from the observation that string
networks can suspend among several parallel D3-branes in such a way that
$1/8$ of the bulk supersymmetries are preserved. This is a
generalization of the interpretation of the 'tHooft-Polyakov monopole as
a D-string suspended between parallel D3-branes in the previous
subsection. 

\begin{figure}[tp]
\begin{center}
\begin{minipage}{12cm}
\begin{center}
\includegraphics[width=7cm]{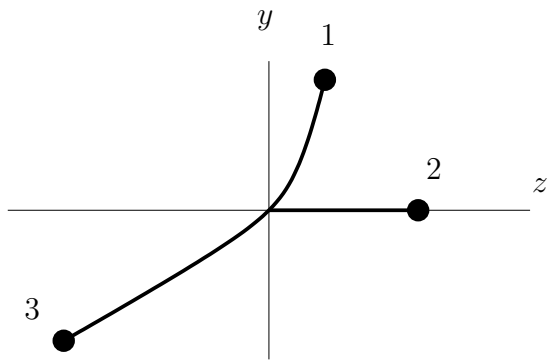}
\put(-104,128){$y$}
\put(-80,120){$1$}
\put(-40,69){$2$}
\put(-192,16){$3$}
\put(0,64){$z$}
\caption{Location of the D3-brane surfaces of a 1/4 BPS dyon solution in
 $SU(3)$, in  course of varying $r$ in the $y$-$z$ space in the bulk
 (solid lines). We have chosen $(\alpha, \beta)=(-1,1/3)$.
 Three blobs are the location of the D3-branes at $r=\infty$. They
 deform their shape to reach the origin $y=z=0$ where the three
 D3-branes meet at $r=0$.} 
\label{figbend}
\end{center}
\end{minipage}
\end{center}
\end{figure}

In \cite{su3}, it was found that the trajectories of the deformed
D3-brane surfaces determined by the eigenvalues of $Y$ and $Z$ in the
solution bend in a nontrivial manner in the bulk spacetime. 
The location of the D3-brane in course of varying $r$ (the radial
direction in the worldvolume) is depicted in Fig.~\ref{figbend}. 
The way they are bent was found \cite{su4}
to be consistent with the ``effective string network''. It is defined
with the ``effective $(p,q)$-strings whose 
$(p,q)$ charges are defined by the electric and magnetic charges
measured inside the sphere of the radius $r$ in the worldvolume. 
The orientation of the effective $(p,q)$-strings are along
the lines tangential to the D3-branes at $r$ in the $Y$-$Z$ plane 
in the bulk. Surprisingly, these effective $(p,q)$-strings
form a stable consistent string network.

Here we evaluate the distribution of D-string and F-string charges in
the bulk directly from the solution given in \cite{su3} to see the
consistency with this picture of ``effective'' string network. 
The solution is spherically symmetric under a linear combination of the 
worldvolume and gauge rotations, 
\begin{eqnarray}
\epsilon_{ijk}x_j\p_k + T_i \quad (i=1,2,3)
\end{eqnarray}
where the first term is a generator of 
the worldvolume rotation, while $T_i$ is a generator of the maximal
embedding of $SU(2)$ in $SU(3)$ with $T_3 = \mbox{diag}(1,0,-1)$.
The solution (rescaled by a factor $2\pi\alpha'$ compared to that of
\cite{su3}) is written on the $x_3$ axis as  
\begin{eqnarray}
&&Y=-2\pi\alpha'\frac{H(r)}{r}T_3 \ ,
\quad 
Z= -A_0 = 2\pi\alpha'\frac{1}{4r}
\mbox{diag}(\varphi_+\!+\!\varphi_-, -2\varphi_-, 
-\varphi_+\!+\!\varphi_-)  
\nn
\end{eqnarray}
where 
\begin{eqnarray}
 \varphi_+ \equiv \alpha H(r) \ , \quad
\varphi_- \equiv \beta
\left(3\coth Pr - Pr \frac{2\cosh^2 Pr + 1}{\sinh^2 Pr}\right) \ ,
\end{eqnarray}
$r\equiv \sqrt{\sum_{i=1}^3(x_i)^2}$, 
and $\alpha, \beta$ are constant parameters.

The D3-brane charge density is found to be 
\begin{eqnarray}
 J_{0123} = 
\sum_{i=1,2,3}\delta(z-\lambda_i(r))\delta(y-\lambda(r)) \ ,
\end{eqnarray}
where $\lambda_i$'s are the eigenvalues of the field $Z$ while 
$\lambda\equiv 2\pi\alpha'H(r)/r$ is for $Y$, 
\begin{eqnarray}
 \lambda_1(r) \equiv \pi\alpha' \frac{\varphi_+ + \varphi_-}{2r}
\ , \quad
 \lambda_2(r) \equiv -\pi\alpha' \frac{\varphi_-}{r}
\ , \quad
 \lambda_3(r) \equiv \pi\alpha' \frac{-\varphi_+ + \varphi_-}{2r}
\ .
\end{eqnarray}
Here $y$ and $z$ are the bulk coordinates corresponding to the
transverse scalars $Y$ and $Z$ respectively.
The location of the D3-branes in the bulk is shown in
Fig.~\ref{figbend}. 
Due to the way of constructing the spherically symmetric solutions,
$Z$ and $Y$ are simultaneously
diagonalizable, and thus the location of the D3-brane is definite in the
bulk. (In the next section, we shall treat the cases where 
two or more transverse scalars are noncommutative with each other.)

Since the solution satisfies the BPS equations 
$2\pi\alpha' B_i = D_i Y$ and $2\pi\alpha' E_i = D_i Z$,
the D-string/F-string charge densities are given in the present case by
\begin{eqnarray}
 \widetilde{J}_{0y} &=& {\cal T}_{\rm D3}\mbox{Str}
\left[\exp\left[ik_z Z + ik_y Y\right]D_iY D_iY\right] \ , 
\nn\\
 \widetilde{J}_{0z} = \widetilde{I}_{0y} 
&=& {\cal T}_{\rm D3}\mbox{Str}
\left[\exp\left[ik_z Z + ik_y Y\right]D_iZ D_iY\right] \ , 
\nn\\
 \widetilde{I}_{0z} 
&=& {\cal T}_{\rm D3}\mbox{Str}
\left[\exp\left[ik_z Z + ik_y Y\right]D_iZ D_iZ\right] \ .
\nn
\end{eqnarray}
A straightforward but tedious calculations show
\begin{eqnarray}
 J_{0y}/{\cal T}_{\rm D3} &=& 
(\lambda')^2\left[
\delta(z-\lambda_1)\delta(y-\lambda)
-\delta(z-\lambda_3)\delta(y+\lambda)
\right]
\nn\\
& &
+ \lambda \frac{2 K^2(r)}{r^2}
\left[
(\theta(y-\lambda)-\theta(y))
\delta\left(
\lambda_2 -z + y\frac{\lambda_1\! -\! \lambda_2}{\lambda}
\right)
\right.
\nn\\
& &\hspace{20mm} 
+
\left.
(\theta(y+\lambda)-\theta(y))
\delta\left(
-\lambda_2 +z + y\frac{\lambda_3\! -\! \lambda_2}{\lambda}
\right)
\right] \ , 
\nn
\\
J_{0z}/{\cal T}_{\rm D3} &=&  I_{0y}/{\cal T}_{\rm D3} \nn \\
& =&  
\lambda' \left[
\lambda_1'\delta(z-\lambda_1)\delta(y-\lambda)
-\lambda_3'\delta(z-\lambda_3)\delta(y+\lambda)
\right]
\nn\\
& &
+  \frac{2 K^2(r)}{r^2}
\left[
(\lambda_1-\lambda_2)(\theta(y-\lambda)-\theta(y))
\delta\left(
\lambda_2 -z + y\frac{\lambda_1\! -\! \lambda_2}{\lambda}
\right)
\right.
\nn\\
& &\hspace{20mm} 
+
\left.
(\lambda_2-\lambda_3)(\theta(y+\lambda)-\theta(y))
\delta\left(
-\lambda_2 +z + y\frac{\lambda_3\! -\! \lambda_2}{\lambda}
\right)
\right] \ , 
\nn
\end{eqnarray}
\begin{eqnarray}
 I_{0z}/{\cal T}_{\rm D3}  &=& 
(\lambda_1')^2\delta(z-\lambda_1)\delta(y-\lambda)
+ (\lambda_2')^2\delta(z-\lambda_2)\delta(y)
+(\lambda_3')^2\delta(z-\lambda_3)\delta(y+\lambda)
\nn\\
& &
+  \frac{2 K^2(r)}{r^2}
\left[
\frac{(\lambda_1-\lambda_2)^2}{\lambda}
(\theta(y-\lambda)-\theta(y))
\delta\left(
\lambda_2 -z + y\frac{\lambda_1\! -\! \lambda_2}{\lambda}
\right)
\right.
\nn\\
& &\hspace{20mm} 
+
\left.
\frac{(\lambda_2-\lambda_3)^2}{\lambda}
(\theta(y+\lambda)-\theta(y))
\delta\left(
-\lambda_2 +z + y\frac{\lambda_3\! -\! \lambda_2}{\lambda}
\right)
\right] \ . \nn
\end{eqnarray}
We have checked that these components of the currents satisfy the
current conservation condition, $\p_z J_{0z} + \p_y J_{0y}=0$, 
$\p_z I_{0z} + \p_y I_{0y}=0$. The fact we can find from these explicit
expressions of the string current is that there are two $(p,q)$-strings
suspended between the D3-branes, one is between the first and the second 
D3-brane located at $(z,y)=(\lambda_1, \lambda)$ and $(\lambda_2,0)$,
another is between the second and the third, 
$(\lambda_2,0)$ and $(\lambda_3, -\lambda)$. These strings take a part
of the 
electro-magnetic charges on the D3-branes out into the bulk, and change
the shape of the D3-branes nontrivially. In fact, one can see that the
first and the third D3-branes are pulled by these bulk strings and
resultantly bend consistently in Fig.~\ref{figbend}. 

\begin{figure}[tp]
\begin{center}
\begin{minipage}{14cm}
\begin{center}
\includegraphics[width=14cm]{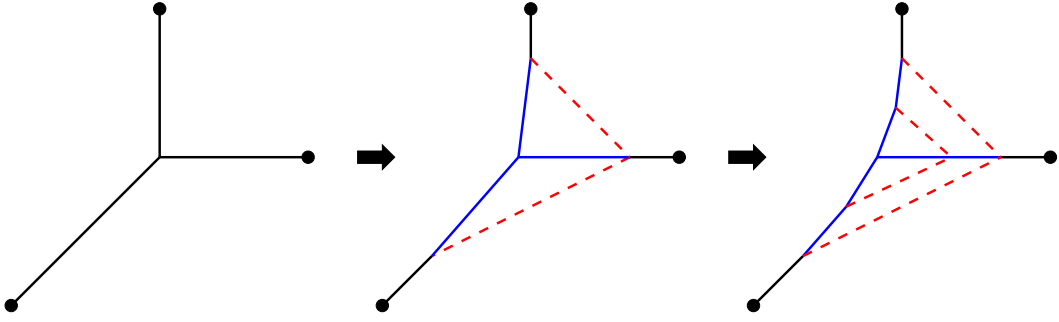}
\caption{We show that the ``effective string network'' found in 
 \cite{su4} is consistent with the bulk $(p,q)$-strings calculated here.
 At $r=\infty$, the lines tangential to the D3-brane (blobs)
 meet at a point (Left), which is an effective string network. 
 The presence of $(p,q)$ strings (dashed lines) 
 in the bulk connecting two of three D3-branes make the D3-brane (solid
 lines) bend in  the $Z$-$Y$  plane (Right), in such a way that at each
 $r$ the effective string network makes sense. 
}
\label{f:prongion}
\end{center}
\end{minipage}
\end{center}
\end{figure}

The ``effective'' string network at $r$ in \cite{su4} was defined as a
collection of ``effective'' $(p,q)$-strings whose charges are provided
by the electric and magnetic charges integrated over a ball whose radius
is $r$. Therefore, actual bulk $(p,q)$ strings should appear as a
difference between the effective string network at $r+\delta r$ and that
at $r$. This is what we found, the strings connecting the first and the
second, and the second and the third, D3-branes.
A graphical description of this charge
conservation is shown in Fig.~\ref{f:prongion}. 

Finally, the energy distribution in the bulk is given by 
\begin{eqnarray}
 T_{00}= J_{0123} + I_{0z} + J_{0y} \ .
\end{eqnarray}
One can obtain similar expression for pressures and check that 
the energy-momentum tensor is conserved, and forces are balanced in the
bulk. 

%%%%%%%%%%%%%%%%%%%%%%%%%%%%%%%%%%%%%%%%%%%%%%%%%%%%%%%%%%%%%%%%%%%%%
%%%%%%%%%%%%%%%%%%%%%%%%%%%%%%%%%%%%%%%%%%%%%%%%%%%%%%%%%%%%%%%%%%%%%
%%%%%%%%%%%%%%%%%%%%%%%%%%%%%%%%%%%%%%%%%%%%%%%%%%%%%%%%%%%%%%%%%%%%%

\section{Fuzzy $S^2$}
\label{sec3}
\setcounter{footnote}{0}

So far, we have studied the D-brane/F-string charge distribution of 
nonabelian worldvolume configurations with a single transverse scalar
fields. (The last example in section \ref{1/4} uses two scalar fields
but they are mutually commutative.) Here and in the following sections,
we go into the examples with more intrinsic noncommutativity -- two or
more scalar fields are mutually noncommutative. We shall see 
how the nontrivial commutation relations among transverse scalar fields
in D0-brane effective field theory give extended distribution of the
charges in the bulk, thorough the supergravity charge density formulas.

In this section, we work with fuzzy two-sphere configuration of 
the matrix scalar fields in the D0-brane low energy effective theory.
Spherical membranes in Matrix theory are known as one of the
fundamental bound states formed by D0-branes \cite{Kabat-Taylor}. 
Explicit configuration of the D0-brane matrices are 
given as 
\begin{eqnarray}
 X_i = \frac{2 R}{N} L_i
\label{solD0s2}
\end{eqnarray}
where $L_i$ ($i=1,2,3$) are an $N$-dimensional 
irreducible representation of generators of $SU(2)$ satisfying
$[L_i, L_j]=i\epsilon_{ijk} L_k$,
so that the matrices $X_i$ are subject to the algebra
\begin{eqnarray}
 [X_i, X_j]=\frac{2R}{N}i\epsilon_{ijk} X_k \ .
\label{su2x}
\end{eqnarray}
This matrix configuration is called a 
fuzzy $S^2$. This radius $R$ can be seen in the relation 
\begin{eqnarray}
 \sum_{i=1}^3 X_i^2 = R^2 \left(1-\frac1{N^2}\right)\identity_N \ ,
\label{locus}
\end{eqnarray}
in which there appears a $1/N^2$ correction. One can absorb this
correction factor into the redefinition of $R$, but we leave it as it
is for simplicity of the calculations. 

This locus equation (\ref{locus})
suggests that the D0-branes are distributed on a
single $S^2$ uniformly. This expectation further leads us to a
conjecture that the fuzzy sphere configuration defined by
(\ref{solD0s2}) is a dual (equivalent another) description of a
spherical D2-brane on which $N$ D0-branes are uniformly bound. 
This conjecture is a generalization of the fact observed in 
section \ref{sec2}
for the fuzzy plane. In the following subsections we shall see
if this conjecture might come out to be true or not, by looking at
various D-brane charge densities generated by the matrix configuration
(\ref{solD0s2}). 

\subsection{D-brane charge distribution}

\subsubsection{D0-brane charge}
\label{3.1}

Let us evaluate the D0-brane charge density (\ref{dens0})
for this fuzzy $S^2$ configuration of D0-branes (\ref{solD0s2}).
In the evaluation we may bring the direction of the vector
$\vec{k}$ along $x_3$, without losing generality. 
This is due to the $SU(2)\sim SO(3)$ symmetry of the D0-brane
configurations (\ref{solD0s2}): 
for any given rotation matrix $R$ of $SO(3)$, there is
an element of $SU(2)$ which satisfies
\begin{eqnarray}
R_{ij}X_j =U(R) X_i U(R)^\dagger. 
\label{rot}
\end{eqnarray}
Then the charge density becomes 
\begin{eqnarray}
 \widetilde{J}_0(\vec{k}) = {\cal T}_{\rm D0}\;
{\rm tr} \left[e^{ikX_3}\right], 
\label{k3}
\end{eqnarray}
where $k \equiv \sqrt{k_1^2 + k_2^2 + k_3^2}$. Hereafter we omit trivial
dependence on the other spatial directions $x^\mu$ with
$\mu=4,5,\cdots,9$. 
We may use a particular representation of the $SU(2)$ generators 
in which the representation of $L_3$ becomes diagonal, 
\begin{eqnarray}
 L_3 = {\rm diag} \left(
\frac{N-1}2, \frac{N-3}2, \cdots, -\frac{N-1}2
\right).
\label{l3}
\end{eqnarray}
With this representation the density (\ref{k3}) is evaluated as
\begin{eqnarray}
 \widetilde{J}_0(\vec{k}) = 
{\cal T}_{\rm D0}\!\!\!\!\!\!\!\!\!\!
\sum_{m=-(N-1)/2}^{(N-1)/2} \!\!\!\!\!\!\!\!\!
\exp\left[
ikm\frac{2R}{N}
\right]=
\left\{
\begin{array}{ll}
{\cal T}_{\rm D0}\!
\displaystyle
\sum_{n=1}^{N/2}\left(e^{ikR(2n-1)/N}\!+\!e^{-ikR(2n-1)/N}\right)
& (\mbox{$N$ even}) \\[3mm]
{\cal T}_{\rm D0}\left(\! 1+\!\!\!\!\!\!
\displaystyle\sum_{n=1}^{(N-1)/2}
\!\!\!\!\left(e^{2ikRn/N}+e^{-2ikRn/N}\right)\!
\right)
& (\mbox{$N$ odd}) 
\end{array}
\right.
\label{D0mom}
\end{eqnarray}
Here the summation for $m$ is with a unit spacing. 

To obtain the inverse Fourier transform of this expression, it is
useful to find the Fourier transform of a delta function 
$\delta(r-a)$ where $r \equiv \sqrt{x_1^2 + x_2^2 + x_3^2}$,
\begin{eqnarray}
 \int\! d^3x\; e^{ik_ix_i} \delta(r-a) 
&=& 
\frac{-2\pi ia}{k}(e^{ika}-e^{-ika}) \ ,
\label{deltar}
\end{eqnarray}
and hence for $\delta'(r-a)$ 
\begin{eqnarray}
 \int\! d^3x\; e^{ik_ix_i} \delta'(r-a) 
&=& -\frac{\p}{\p a} \int\! d^3x\; e^{ik_ix_i} \delta(r-a) 
\nn\\
&=&
 \frac{2\pi i}{k}(e^{ika}-e^{-ika}) 
-2\pi a (e^{ika}+e^{-ika}) \ . 
\end{eqnarray}
{}From these, we obtain a relevant formula for the inverse 
Fourier transform in three dimensions,
\begin{eqnarray}
D(r,a)\equiv\frac{-1}{2\pi a}
\left(
\frac{1}{a}\delta(r-a)+\delta'(r-a)
\right) 
\quad \mathop{\longrightarrow}^{\mbox{\tiny F.T.}} \quad 
e^{iak}+ e^{-iak} \ .
\label{ftformula}
\end{eqnarray}
Using this formula, the inverse Fourier transform
of the D0-brane density (\ref{D0mom}) is given by
\begin{eqnarray}
 J_0(x) = 
\left\{
\begin{array}{ll}
{\cal T}_{\rm D0}
\sum_{n=1}^{N/2}D(r,R^{(n)})
& (\mbox{$N$ even}) \\[3mm]
{\cal T}_{\rm D0}\left(\delta^3(x) + 
\sum_{n=1}^{(N-1)/2}D(r,R^{(n)})
\right)
& (\mbox{$N$ odd}) 
\end{array}
\right.
\label{d0n}
\end{eqnarray}
where $D(r,R^{(n)})$ is a spherical delta function (plus its derivative)
with a support on a spherical shell whose radius $R^{(n)}$ 
($n=1,2,\cdots,[N/2]$) is given by
\begin{eqnarray}
R^{(n)}  \equiv 
\left\{\begin{array}{lc}
\frac{2n-1}{N}R 
& (\mbox{$N$ even}) \\[3mm]
\frac{2n}{N}R 
& (\mbox{$N$ odd}) 
\end{array}
\right.
\label{defrn}
\end{eqnarray}
This means that the D0-branes are distributed on a collection of 
two-spheres (shells) with these different radii. 
Since we get an integration of each shell as 
\begin{eqnarray}
 \int\! 4\pi r^2 dr \; D(r,R^{(n)}) = 2 \ ,
\end{eqnarray}
we find that each shell is composed of two D0-branes. 
When $N$ is odd, the remaining single 
D0-brane is located at the origin.

The result that the D0-branes are distributed on a collection of
shells seems quite strange.\footnote{M. van Raamsdonk has considered a
related puzzle \cite{mark}.} 
Naively one expects from the matrix equation
showing the spherical locus (\ref{locus}) that all $N$ D0-branes would
have been distributed on a single spherical shell with the radius $R$
corrected by $1/N^2$. The fuzzy sphere, a bound state of D0-branes, 
given by (\ref{solD0s2}) is expected to have a dual description by a
single spherical D2-brane with magnetic field on it.
However, the direct calculation of the D0-brane charge distribution
shows that there are many spherical shells with various radii, on the
contrary, and our naive expectation is not the case.
Therefore, with use of the charge density formulas adopted in
this paper, the duality for finite $N$ does not seem to hold. 
In the next subsection we shall see that this discrepancy is resolved if
we take the continuum (large $N$) limit.

For a later convenience, here let us see a consistency of 
our computation. 
When $k_1=k_2=0$, the D0-brane density for the fuzzy $S^2$ is given by
\begin{eqnarray}
 \widetilde{J}_0(k_1=k_2=0, k_3) =
{\cal T}_{\rm D0}\!\!\!
\sum_{m=-(N-1)/2}^{(N-1)/2} 
\exp\left[
ik_3m\frac{2R}{N}
\right]
\end{eqnarray}
And the Fourier transform of this is given by
\begin{eqnarray}
 \int\!dx_1 dx_2 \;J_0(\vec{x}) = {\cal T}_{\rm D0}\!\!\!\! 
\sum_{m=-(N-1)/2}^{(N-1)/2} \!\!\!\! 
\delta(z-2mR/N).
\end{eqnarray}
This seems to be contradictory to 
the result (\ref{d0n}) since each shell consists of two D0-branes.
Normally one may expect
that for a uniformly distributed D0-branes on a shell, an integration
over two dimensions orthogonal to the $x_3$ axis gives a uniform
distribution of D0-branes on the axis, not the delta functions at the
poles of the sphere. 
This apparent discrepancy can be resolved if one notes the second term
in the density function $D(r,R^{(n)})$ (\ref{ftformula}). In fact, 
using polar coordinates, we can calculate for $N=2$ the relevant
quantity as 
\begin{eqnarray}
 \int \! dx_1 dx_2 \; J_0(\vec{x})
&=&
\int\! 2\pi \rho d\rho \; J_0(\vec{x})
\nn\\
&=&
\frac{- {\cal T}_{\rm D0}}{ (R/2)^2} \int\! \rho d \rho \; 
\left(\delta(r-R/2) + \frac{R}{2}\delta'(r-R/2)\right)
\nn\\
&=& {\cal T}_{\rm D0} \left(\delta(x_3-R/2) + \delta(x_3+R/2)\right),
\label{strange}
\end{eqnarray}
where $\rho\equiv \sqrt{x_1^2+x_2^2}$.
So the strange term of the derivative of the delta function works for
consistency.

\subsubsection{D2-brane charge}
\label{3.2}

To see what is a possible dual D2-brane description might be, let us
evaluate the D2-brane charge density (\ref{dens2})
generated by the fuzzy $S^2$ configuration (\ref{solD0s2}). 
We note that from the rotational covariance of the formula (\ref{dens2})
the final expression of this charge density should be of the form  
$\widetilde{J}_{0ij}= k_m \epsilon_{ijm} f(k)$ where $f(k)$ is 
some function of the total momentum $k$.
Based on this fact, we may 
put $k_1=k_2=0$ for simplicity and later we can
recover all the $k_i$  dependence from the covariance. 
For the momentum along $x_3$ direction, we have
\begin{eqnarray}
 \widetilde{J}_{012}(k_1=k_2=0, k_3) & = & 
\frac{{\cal T}_{\rm D0}}{2\pi\alpha'}\left(\frac{2R}{N}\right)^2
{\rm tr}\left[
L_3 e^{ik(2R/N)k_3 L_3}
\right] 
\nn\\
& = &
\frac{{\cal T}_{\rm D0}}{2\pi\alpha'}\left(\frac{2R}{N}\right)^2
\sum_{m=-(N-1)/2}^{(N-1)/2}m \; e^{imk_3(2R/N)}.
\end{eqnarray}
So from the rotational covariance we obtain
\begin{eqnarray}
 \widetilde{J}_{0ij}(\vec{k})=
\frac{{\cal T}_{\rm D0}}{2\pi\alpha'}\left(\frac{2R}{N}\right)^2
\sum_{m=-(N-1)/2}^{(N-1)/2} m\frac{\epsilon_{ijl}k_l}{k} 
\; e^{imk(2R/N)}.
\end{eqnarray}
Using the formula for an inverse Fourier transform
\begin{eqnarray}
 \frac{\epsilon_{ijl}k_l}{k} (e^{ika}-e^{-ika})
\quad \mathop{\longleftarrow}^{\mbox{\tiny F.T.}} \quad 
-\frac{1}{2\pi a}\frac{\epsilon_{ijl}x_l}{r}\delta'(r-a) 
\end{eqnarray}
and 
${\cal T}_{\rm D2}={\cal T}_{\rm D0}/(4\pi^2\alpha')$, we obtain
\begin{eqnarray}
 J_{0ij}(\vec{x})&=&
-{\cal T}_{\rm D2}\left(\frac{2R}{N}\right)
\epsilon_{ijk}\frac{x_k}{r}
\sum_{n=1}^{[N/2]} \delta'(r-R^{(n)}).
\label{d2n}
\end{eqnarray}
This shows that locally the D2-brane charge is induced on the shells 
of the radii $R^{(n)}$.

Moreover, it is found that the induced D2-brane charge on a single shell
is fractional. To see this, we extract only the $n$-th 
shell and compute its total charge as 
\begin{eqnarray}
 \int\!
 d^3x\; \frac12  \epsilon_{ijk} \hat{x}_k J_{0ij}^{(n)}(\vec{x})
&=& 
-\int \!4\pi r^2 dr \; {\cal T}_{\rm D2} \left(\frac{2R}{N}\right)
\delta'(r-R^{(n)})
\nn\\
&=&
8\pi R^{(n)} {\cal T}_{\rm D2}\left(\frac{2R}{N}\right)
\nn\\
&=&
{\cal T}_{\rm D2}4\pi (R^{(n)})^2 \cdot 
\left(\frac{4R}{NR^{(n)}}\right).
\end{eqnarray}
Since the factor $4\pi (R^{(n)})^2$ is the area of the shell, 
we may conclude that the $n$-th shell has only a fraction 
$4/(2n-1)$ (or a fraction $2/n$) of a D2-brane for even (or odd) $N$.

We have encountered two strange results : D2-brane charges are induced
on the shells with various radii as is the case for the D0-brane charge
density, and furthermore, each shell has a fractional D2-brane charge. 
This strange distribution disappear in the large $N$ limit, as we will
see in the next subsection.

\subsection{Large $N$ limit to the continuum}
\label{3.3}

The results obtained in the previous subsections seem 
rather strange, because one might expect that $N$ D0-branes 
in the configuration (\ref{solD0s2}) form a single $S^2$ 
which is a conventional interpretation consistent with a dual D2-brane
picture of a spherical brane with a magnetic field.
However, in the sense of matrix regularization of membranes, 
at least if we take a large $N$ limit with fixed $R$, 
we might be able to have this dual correspondence and 
see the exact relation between the two. 
The dual D2-brane picture is based on a continuous worldvolume of
D2-brane which would be reproduced by the continuum large $N$ limit.
(Note that for the democratic duality we expect, the D-brane charge
should be anyway found on a single shell even for finite $N$. We propose
a resolution of this problem in section \ref{sec6}.)

So for the present let us follow this idea of the matrix regularization
and take the large $N$ limit carefully. 
We will see that although there are many shells for finite $N$,
in the large $N$ limit the shells annihilate with each other and only 
the most outer shell remains, to give a consistent dual D2-brane
picture. 

First we examine the D2-brane charge density (\ref{d2n}). In the large
$N$ limit, the summation over $n$ in (\ref{d2n}) can be replaced by an
integration over 
a new variable $y$ with the following correspondence
\begin{eqnarray}
y \equiv  \frac{2n}{N}R \ ,
\label{defy}
\end{eqnarray}
for odd $N$ for simplicity.
Then (\ref{d2n}) becomes 
\begin{eqnarray}
 J_{0ij}(\vec{x})& = &
\lim_{N \rightarrow \infty}
-{\cal T}_{\rm D2}
\epsilon_{ijk}\frac{x_k}{r}
\sum_{m>0}^{(N-1)/2} \left(\frac{2R}{N}\right)
\delta'(r-2mR/N)
\nn\\
&=&
-{\cal T}_{\rm D2}
\epsilon_{ijk}\frac{x_k}{r}
\int_0^R\! dy \;\delta'(r-y)
\nn\\
&=&
{\cal T}_{\rm D2}
\epsilon_{ijk}\frac{x_k}{r}
\left[\delta(r-R)-\delta(r)\right].
\end{eqnarray}
Hence the D2-brane charge is found only on a single shell with the
radius $R$, as mentioned. 
The second term $\delta(r)$ appears due to the ill-definedness of
the spherical coordinates, and in fact in the total D2-brane charge it
does not give any contribution due to the Jacobian $4\pi r^2$ of the
spherical coordinates, 
\begin{eqnarray}
 \int \! 4\pi r^2 dr \; 
\frac12  \epsilon_{ijk}\hat{x}_k J_{0ij} 
=  4\pi R^2 {\cal T}_{\rm D2}.
\end{eqnarray}
Thus we have found that, in the large $N$ limit, 
the charge density formula given in \cite{WT-Mark}
provides expected distribution of D2-brane charge 
for the fuzzy $S^2$ configuration of D0-branes (\ref{solD0s2}).

Next, let us check the large $N$ limit of the 
D0-brane density (\ref{d0n}). The large $N$ limit
should give a diverging result since
this $N$ represents the number of the total D0-branes. Thus we have to
divide the expression (\ref{d0n}) by $N$ so that we can see the  
distribution ratio of the component D0-branes. 
\begin{eqnarray}
 \lim_{N\rightarrow \infty}\frac{1}{N}J_0(x)
& = &  
\frac{-{\cal T}_{\rm D0}}{4\pi R}\int_0^R \! dy \;
\frac{1}{y} \left(
\frac{1}{y}\delta(r-y) + \frac{\p}{\p r}\delta(r-y)
\right)
\nn\\
&=&
\frac{-{\cal T}_{\rm D0}}{4\pi R}
\left[
\int_0^R \! dy \;
\frac{1}{y^2} \delta(r-y) 
+\left[\frac{-1}{y}\delta(r-y)\right]_{y=0}^R
-\int_0^R \! dy \;
\frac{1}{y^2}\delta(r-y)
\right]
\nn\\
&=&
\frac{{\cal T}_{\rm D0}}{4\pi R^2}\delta(r-R) +
\frac{{\cal T}_{\rm D0}}{4\pi R}
\lim_{y\rightarrow 0}\frac{-1}{y}\delta(r-y) \ .
\end{eqnarray}
The second term is again an ill-defined quantity due to the spherical
coordinates, and vanishes when we integrate over all the bulk space, 
as we shall see below. So only the
first term is of physical interest, and this shows that in the large $N$
limit all the D0-branes are localized on the single sphere whose radius
is $R$. A consistency check of the total number of D0-branes is 
\begin{eqnarray}
\int \!\! 4\pi r^2 dr
  \lim_{N\rightarrow \infty}\frac{1}{N}J_0(x)
= \int \!\! 4\pi r^2 dr
\left(
\frac{{\cal T}_{\rm D0}}{4\pi R^2}\delta(r\!-\!R) +
\frac{{\cal T}_{\rm D0}}{4\pi R}
\lim_{y\rightarrow 0}\frac{-1}{y}\delta(r\!-\!y)
\right)
={\cal T}_{\rm D0}.
\nn
\end{eqnarray}

We conclude that in the large $N$ limit the fuzzy $S^2$ configuration 
(\ref{solD0s2}) provides a single spherical D2-brane with $N$ D0-branes 
on it. The radius of this sphere is consistent with the spherical locus
relation (\ref{locus}). From this observation, the strange appearance of
the many shells for the finite $N$ case can be understood as a kind of   
$1/N$ corrections, in a sense. It was already found in
\cite{Kabat-Taylor} that the configuration (\ref{solD0s2}) generates
correct large distance behaviour of the gravitational field up to $1/N$
corrections, which night be consistent with our results.

\subsection{Another large $N$ limit to the fuzzy plane}
\label{3.4}

We may consider varying also $R$ as $N$ increases, and this makes it
possible to relate the fuzzy sphere with the fuzzy (noncommutative)
plane studied in subsection \ref{ncp}. Basically this limit is defined
as enlarging the north pole of the fuzzy sphere by taking 
$R \rightarrow \infty$ while keeping a finite noncommutativity on the
sphere \cite{limit}. 
More precisely, We first define a new $X_3$ by a shift from the
original one, 
\begin{eqnarray}
 X_3 \equiv X_3^{\rm original} - R \identity_N \ ,
\end{eqnarray} 
so that the large $R$ limit does not bring the north pole to a spatial
infinity. Then the algebra of the fuzzy sphere becomes 
\begin{eqnarray}
  [X_1, X_2]=\frac{2R}{N}i X_3  + \frac{2iR^2}{N} \identity_N \ ,
\label{anoal}
\end{eqnarray}
while the other commutators are the same as before, (\ref{su2x}).
We take a large $N$ limit while keeping $2R^2/N \equiv \theta$
fixed, then the fuzzy sphere algebra (\ref{su2x}) and (\ref{anoal})
reduces to the fuzzy plane algebra (\ref{ncplane}). 

Let us check if this limit of the results in subsections \ref{3.1} and
\ref{3.2} reproduces those of subsection \ref{ncp}. For finite $N$, 
the D0-brane and D2-brane charges are the same as (\ref{d0n})
and (\ref{d2n}) except the shift in the $x_3$ direction, 
$x_3 \rightarrow x_3 +R$. Since we fix $\theta$, the radius 
$R \sim \sqrt{N}$ goes to the infinity, so we may expand $r$ as 
\begin{eqnarray}
 r = \sqrt{(x_1)^2+ (x_2)^2 + (x_3 + R)^2}
= R \left(1 + \frac{x_3}{R} 
+ {\cal O}\left(\frac{1}{R^2}\right)\right) \ .
\end{eqnarray}
With this in mind, we may renumber the shell as $2(n'-1)= N-2n$
($n'=1,2, \cdots, N/2$) and find the D0-brane charge distribution
(\ref{d0n}) as 
\begin{eqnarray}
&& J_0(x) = {\cal T}_{\rm D0}\sum_{n'=1}^{N/2}
\frac{-1}{2\pi\left(R-\frac{2n'-1}{N}R\right)}
\left(
\frac{1}{R\!-\!\frac{2n'\!-\!1}{N}R}
\delta\!\left(\!x_3\! +\! \frac{2n'\!-\!1}{N}R\!\right)
\right.
\nn\\
&&\hspace{70mm}
\left.
+ \delta'\!\left(\!x_3\! +\! \frac{2n'\!-\!1}{N}R\!
\right)
+ \mbox{higher}
\right) \ .
\end{eqnarray}
Introducing a new variable $y' \equiv \frac{2n'-1}{N}R$, we find 
the measure of
the integration is given as $dy = 2R/N = \theta/R$, and so
\begin{eqnarray}
\lefteqn{\lim_{N\rightarrow \infty} J_0(x)}
\nn\\ &=&
\lim_{N\rightarrow \infty} {\cal T}_{\rm D0}
\sum_m \frac{-1}{2\pi(R-y)}\left(\frac{1}{R-y}\delta(x_3+y)
+ \delta'(x_3+y)\right) \nn \\
& = &
\lim_{N\rightarrow \infty} {\cal T}_{\rm D0}
\sum_m \frac{-1}{2\pi R}\left(1\!+\!\frac{y}{R}+ \cdots\right)
\left(\frac{1}{R}\left(1\!+\!\frac{y}{R}+ \cdots\right)\delta(x_3+y)
+ \delta'(x_3+y)\right) \nn \\
&=& 
{\cal T}_{\rm D0} \int_0^\infty 
\frac{-1}{2\pi} \frac{dy}{\theta} \delta'(x_3+y)
\nn\\
&=& \frac{{\cal T}_{\rm D0}}{2\pi\theta} \delta(x_3).
\end{eqnarray}
In the last equality we discard the delta function at the infinity. 
This last expression coincides with (\ref{ncp1}), as expected.
Here, the shells disappear due to the cancellation with each other as in
the previous subsection. So, even though the algebra in the large $N$
limit is still nontrivial, the limit makes the cancellation occur and we
have finally a good agreement with the dual D2-brane picture. 
The results in this subsection might lead us to expect that the large
$N$ limit is necessary to have a result consistent with the dual
D2-brane picture. 
However, we shall see in section \ref{sec5} that one of the
other kinds of the large $N$ limits is not the case.

\section{Fuzzy $S^4$}
\label{sec4}

It is straightforward 
to generalize the computation in the previous section to a fuzzy $S^4$.
An explicit expression for the fuzzy $S^4$ constructed from D0-branes
was provided in \cite{Castelino} as
\begin{eqnarray}
 X_i = \frac{R}{n} G_i^{(n)} \quad (i=1,2,\cdots,5)
\label{fuzzys4sol}
\end{eqnarray}
where $R$ is an expected radius of the sphere, and 
\begin{eqnarray}
 G_i^{(n)}\equiv
\left[
(\Gamma_i\otimes \identity \otimes \identity \cdots)
+ (\identity \otimes \Gamma_i \otimes \identity \cdots )+ \cdots
\right]_{\mbox{sym}} ,
\label{solD0s4}
\end{eqnarray}
which is an $n$-fold symmetric tensor product of the following matrices
\begin{eqnarray}
&& \Gamma_a = 
\left(
\begin{array}{cc}
0 & -i\sigma_a \\
i \sigma_a & 0
\end{array}
\right) \; (a=1,2,3) \  , 
\quad
 \Gamma_4 = 
\left(
\begin{array}{cc}
0 & \identity_2 \\ \identity_2 & 0
\end{array}\right) 
\ ,
\quad 
\Gamma_5= 
\left(
\begin{array}{cc}
\identity_2 & 0 \\ 0 & -\identity_2
\end{array}\right) \ .
\hspace{5mm}
\end{eqnarray}
The rank of matrices (the total number of D0-branes) $N$ is related to
this $n$ as 
\begin{eqnarray}
 N=\frac16(n+1)(n+2)(n+3) \ .
\end{eqnarray}
We have a relation analogous to (\ref{locus}) in the fuzzy $S^2$,
\begin{eqnarray}
 \sum_{i=1}^{5} (X_i)^2 = R^2 \left(1+ \frac4n\right)\identity_N \ ,
\label{4locus}
\end{eqnarray}
which indicates that the D0-branes are on a
spherical locus of the radius $R$ corrected by $1/n$. 
We will see in this section that 
the large $N$ limit of the D0-brane configuration (\ref{solD0s4}) is
actually representing a bound state of $n$ coincident
spherical D4-brane and $N$ uniformly distributed D0-branes, forming 
a four-sphere shell of radius $R$.

\subsection{D-brane charges}
\label{4.2}

\subsubsection{D0-brane charge}
\label{4.2.1}

Due to the rotational invariance $SO(5)$ of the 
above D0-brane configuration explained in \cite{Castelino}, 
which is analogous to (\ref{rot}) for the fuzzy $S^2$, 
we can bring the vector $\vec{k}$
along the $x^5$ axis and evaluate the D0-brane charge density,
as in the case of the previous section.
We obtain
\begin{eqnarray}
 \widetilde{J}_0(\vec{k})= {\cal T}_{\rm D0}
{\rm tr}
\left[
e^{ik(R/n)G_5^{(n)}}
\right] \ ,
\end{eqnarray}
where in this section $k\equiv \sqrt{\sum_{i=1}^5 k_i^2}$.
In  \cite{Castelino}, it was shown that the number of the eigen vectors
with the eigenvalue $m$ of $G_5^{(n)}$ is $((n+2)^2-m^2)/4$. Using this 
result, we obtain
\begin{eqnarray}
  \widetilde{J}_0(\vec{k})= {\cal T}_{\rm D0}
\sum_{l=0}^{n}\left(
e^{ik(R/n)(2l-n)}\frac{(n+2)^2-(2l-n)^2}{4}
\right) \ .
\end{eqnarray}
What we need for the inverse Fourier transform
in 5 dimensions is that of $\delta(r-R)$
where $r\equiv \sqrt{x_1^2+ \cdots + x_5^2}$, 
\begin{eqnarray}
 \int\! d^5x \; e^{ik_i x_i} \delta(r-a)
&=& \frac{4\pi^2 a}{(ik)^3} 
\left[
-(e^{ika}-e^{-ika}) + ika (e^{ika} + e^{ika})
\right] \ .
\label{formula}
\end{eqnarray}
{}From this equation and its derivatives, we obtain a formula of the
inverse Fourier transform in 5 dimensions 
\begin{eqnarray}
e^{ika} + e^{-ika} 
\quad \mathop{\longleftarrow}^{\; \mbox{\tiny F.T.}} \quad 
\frac{1}{4\pi^2 a^2}
\left[
\frac{3}{a^2}\delta(r-a) + \frac{3}{a} \delta'(r-a) + \delta''(r-a)
\right]\equiv D^{(4)}(r,a) \ . \;\;
\end{eqnarray}
Using this formula, the D0-brane density is found to be
\begin{eqnarray}
 J_0(\vec{x})= 
\left\{
\begin{array}{ll}
{\cal T}_{\rm D0}
\displaystyle\sum_{s=0}^{(n-1)/2}
\displaystyle
\frac{(n+2)^2-(2s+1)^2}{4} D^{(4)}(r,R^{(s)})
&
(\mbox{odd $n$}) \\
{\cal T}_{\rm D0}\displaystyle
\frac{(n+2)^2}{4}\delta^5(x) + 
{\cal T}_{\rm D0}\displaystyle\sum_{s=1}^{n/2}
\displaystyle
\frac{(n+2)^2-(2s)^2}{4} D^{(4)}(r,R^{(s)})
&
(\mbox{even $n$}) 
\end{array}
\right.
\label{d0s4}
\end{eqnarray}
where 
\begin{eqnarray}
R^{(s)}\equiv 
\left\{
\begin{array}{ll}
\displaystyle
\frac{2s+1}{n}R & \mbox{(odd $n$)} \\
\displaystyle
\frac{2s}{n}R & \mbox{(even $n$)} \\
\end{array}
\right.
\end{eqnarray}
We can see that the D0-branes form a collection of $S^4$ 
shells whose radii are given by $R^{(s)}$. The situation is
similar to what we have found in the case of fuzzy $S^2$ in the previous 
section. 

\subsubsection{D2-brane charge}
\label{4.2.2}

We expect that the D0-brane configuration (\ref{solD0s4}) does not
generate any D2-brane charge but gives only a D4-brane charge. 
The D2-brane charge $\widetilde{J}_{0ij}$ is proportional to
\begin{eqnarray}
 {\rm Str}\left(
[X_i, X_j]e^{ik\cdot X}
\right).
\end{eqnarray}
We make again use of the $SO(5)$ symmetry to bring it to the form
\begin{eqnarray}
 {\rm tr}\left(
G^{(n)}_{ij}e^{ik_5 (R/n)G_5}
\right),
\end{eqnarray}
where we have defined $G_{ij}^{(n)}\equiv [G^{(n)}_i, G^{(n)}_j]/2$.
When $i$ or $j$ is $5$, the trace vanishes. For $i,j\neq 5$, we use
the argument of \cite{Castelino} to show that this vanishes :
the eigenvectors of $G_5$ for a given eigenvalue are split into two
groups, one with positive eigenvalues of $G_{ij}$ and the other with
negative eigenvalues of it, for given $i$ and $j$. The number of them
are the same and so the above trace vanishes. 
This means that there is no D2-brane charge 
induced from the fuzzy $S^4$ configuration of D0-branes (\ref{solD0s4}).

\subsubsection{D4-brane charge}
\label{4.2.3}

Finally let us evaluate the D4-brane charge (\ref{dens4})
which is expected to be generated in a spherical shape.
In the evaluation of, say, a component 
$\widetilde{J}_{01234}$, we encounter the following kind of
traces in the symmetrization : 
\begin{eqnarray}
{\rm tr}\left(G^{(n)}_{12}\left(k_i(R/n)G^{(n)}_i\right)^{t}
G^{(n)}_{34}\left(k_i(R/n)G^{(n)}_i\right)^{s-t}
\right) 
\end{eqnarray}
with a various integer $t$ for a given integer $s$.
We assume that after the symmetrization and the trace this is
equal to\footnote{We have checked this
explicitly for $n=1$.}
\begin{eqnarray}
{\rm tr}\left(G^{(n)}_{12}G^{(n)}_{34}\left(k_i(R/n)G^{(n)}_i\right)^{s}
\right) ,
\end{eqnarray}
which simplifies the following 
calculation, since we may use the relation \cite{Castelino}
\begin{eqnarray}
 \epsilon_{ijkl5}X_i X_j X_k X_l = 8(n+2) X_5 (R/n)^4.
\end{eqnarray}

Using the same argument as before, from the rotational covariance under
$SO(5)$ the charge density should be of the form 
$\widetilde{J}_{01234}\propto k_5f(k)$. Then we may bring $\vec{k}$
to the direction of $x_5$ as 
\begin{eqnarray}
 \widetilde{J}_{01234}(\vec{k})&=& 
\frac{{\cal T}_{\rm D0}}{2(2\pi\alpha')^2}
8(n+2)(R/n)^4
{\rm tr}\left(G_5^{(n)}e^{ik(R/n)G_5^{(n)}}\right)
\frac{k_5}{k}.
\end{eqnarray}
The evaluation of this trace is almost the same as that for the D0-brane 
charge, and gives
\begin{eqnarray}
 {\rm tr}\left(G_5^{(n)}e^{ik(R/n)G_5^{(n)}}\right)
= \sum_{l=0}^{n}e^{ik(R/n)(2l-n)} \frac{(n+2)^2-(2l-n)^2}{4}(2l-n).
\end{eqnarray}
Using a formula analogous to (\ref{formula}), 
we obtain the inverse Fourier transform for odd $n$ for example as 
\begin{eqnarray}
\hspace{-4mm} J_{01234}(x)
&=&\frac{{\cal T}_{\rm D0}}{2(2\pi\alpha')^2}
8(n+2)\left(\frac{R}{n}\right)^4
\frac{x_5}{r}
\nn\\
&&
\hspace{-24mm}
\times\!\!\sum_{s=0}^{(n-1)/2}
\frac{(n\!+\!2)^2\!-\!(2s\!+\!1)^2}{4}(2s\!+\!1)
\frac{1}{4\pi^2(R^{(s)})^2}
\left(
\frac{1}{R^{(s)}} \delta'(r-R^{(s)}) + \delta''(r-R^{(s)})
\right).
\label{d4s4}
\end{eqnarray}
We have again plenty of shells of induced D4-brane charges.
The location of the D4-brane shells are the same as that of
D0-branes (\ref{d0s4}).

\subsection{Large $N$ limit to the continuum}
\label{4.3}

As in the previous section of the fuzzy $S^2$, we shall take the large
$N$ limit and see if the above results are consistent in the continuum
limit with the
expected dual picture that there is a bound state of $n$ coincident 
spherical D4-brane shells in which $N$ D0-branes are bound.
First let us consider the large $N$ limit of the D0-brane charge
(\ref{d0s4}).
In the large $N$ limit, we may replace the discrete sum in (\ref{d0s4})
by an integration over a continuous parameter
\begin{eqnarray}
y \equiv \frac{2s+1}{n}R \ ,
\label{repl}
\end{eqnarray}
which is analogous to the previous (\ref{defy}),  but note that in the
present case the denominator is not $N$ but $n$.
Subsequently, the infinitesimal displacement is $\frac{2R}{n}=dy$,
then the result (\ref{d0s4}) of the D0-brane density becomes
in the large $N$ limit
\begin{eqnarray}
&& \lim_{n\rightarrow \infty}\frac1{N} J_0(x)
=
 \lim_{n\rightarrow \infty}
\frac{6T_{{\rm D0}}}{n}
\sum_{s=0}^{(n-1)/2}
\frac{1-((2s+1)/n)^2}{4}D^{(4)}(r,R^{(s)})
\nn\\
&&\hspace{10mm}=
\frac{3T_{{\rm D0}}}{R}
\int_0^R\!dy\;\frac14 \left(1-\frac{y^2}{R^2}\right)
\frac{1}{4\pi^2y^2}\left[
\frac3{y^2}\delta(r-y) + 
\frac3{y}\delta'(r-y) + 
\delta''(r-y)
\right].
\nn
\end{eqnarray}
Performing partial integrations, we are finally led to the result
\begin{eqnarray}
&& \lim_{n\rightarrow \infty}\frac1{N} J_0(x)
=
\frac{{\cal T}_{\rm D0}}{\Omega_4R^4}\delta(r-R)
\nn\\
&&\hspace{30mm}
+\frac{3{\cal T}_{\rm D0}}{16\pi^2 R^3}\lim_{y\rightarrow 0}
\left[
\left(\frac{R^2}{y^3}-\frac3{y}\right)\delta(r-y)
+ \left(
\frac{R^2}{y^2}-1
\right)\delta'(r-y)
\right].\nn
\end{eqnarray}
The second term is irrelevant (and in fact vanishes when integrated over
the whole space), so we have confirmed that in the large $N$ limit all
the D0-branes are on a single shell of an $S^4$ whose volume is
$\Omega_4R^4$ (where $\Omega_4$ is the volume of a unit four-sphere).

Next, let us evaluate the large $N$ limit of the D4-brane 
charge (\ref{d4s4}). From the observation in \cite{Castelino}, we expect
that $n$ coincident spherical D4-branes are generated. So we divide the
D4-brane charge (\ref{d4s4}) by $n$ and take the large $n$ limit.
Due to the replacement (\ref{repl}), we have 
\begin{eqnarray}
 \lim_{n\rightarrow\infty}\frac1{n}
J_{0ijkl}(x)
=
 \frac{{\cal T}_{\rm D0}R^2}{2(2\pi\alpha')^2}
\epsilon_{ijklm}\frac{x_m}{r}\!\!
\int_0^R\!\!\!\! dy 
\left(1\!-\!\frac{y^2}{R^2}\right)
\frac{1}{4\pi^2}\left(
\frac1{y^2}\delta'(r\!-\!y)+ \frac{1}{y}\delta''(r\!-\!y)
\right).\nn
\end{eqnarray}
After a partial integration, we obtain
\begin{eqnarray}
 \lim_{n\rightarrow\infty}\frac1{n}
J_{0ijkl}
&=&
{\cal T}_{\rm D4}
\epsilon_{ijklm}\frac{x_m}{r}
\left[
(\delta(r-R)-\delta(r))
+ \frac{R^2}{2}\lim_{y\rightarrow 0}\frac1y\delta'(r-y)
\right].
\end{eqnarray}
We have used the relation 
${\cal T}_{\rm D0} = (4\pi^2 \alpha')^2{\cal T}_{\rm D4}$. 
In this result, $\delta(r)$ term and the term with $y\rightarrow 0$ are
artificial due to the ill-definedness of the spherical coordinates
at the origin, and actually it vanishes when integrated over the whole
space as before. So we conclude that in the
large $n$ limit, $n$ spherical D4-branes are generated with radius $R$.
\begin{eqnarray}
 J_{0ijkl} \mathop{\rightarrow}_{n \rightarrow \infty}
n {\cal T}_{\rm D4}
\epsilon_{ijklm}\frac{x_m}{r}
\delta(r-R).
\end{eqnarray}
In fact, the total induced D4-bane charge is evaluated as 
\begin{eqnarray}
\lim_{n\rightarrow \infty} \frac{1}{n}\Omega_4\int\!  r^4 dr\;
 J_{0ijkl}\hat{x}_m\epsilon^{ijklm}\frac{1}{4!}
= \Omega_4 R^4 {\cal T}_{\rm D4}.
\end{eqnarray}

%%%%%%%%%%%%%%%%%%%%%%%%%%%%%%%%%%%%%%%%%%%%%%%%%%%%%%%%%%%%%%%%%%%%%
%%%%%%%%%%%%%%%%%%%%%%%%%%%%%%%%%%%%%%%%%%%%%%%%%%%%%%%%%%%%%%%%%%%%%
%%%%%%%%%%%%%%%%%%%%%%%%%%%%%%%%%%%%%%%%%%%%%%%%%%%%%%%%%%%%%%%%%%%%%

\section{Fuzzy cylinder}
\label{sec5}

Among various fuzzy objects, one of the interesting
configurations is a fuzzy cylinder \cite{BakLee,cyl} whose shape is
nontrivial. The object itself is non-compact and elongated to spacetime
infinities, while the fuzzy spheres considered in the previous sections 
are compact objects. In this section we demonstrate similar
calculations of the RR charge distributions for the fuzzy cylinder.
The difference from the results of section \ref{sec3} and \ref{sec4} is
that the 
representation of the fuzzy cylinder algebra is infinite dimensional, so
the matrix is already in large $N$, but the algebra is still
nontrivial and hence not in the continuum limit.

The fuzzy cylinder is defined as a representation of the following
algebra 
\begin{eqnarray}
 [X_1, X_2]=0, \quad 
 [X_2, X_3]=i \rho_3 X_1, \quad 
 [X_1, X_3]=-i \rho_3 X_2 \ .
\label{cyal}
\end{eqnarray}
This algebra can be obtained by a certain deformation of the fuzzy $S^2$
algebra (\ref{su2x}). This deformation will be studied in subsection
\ref{5.4}. One of the explicit representations which we shall use in
this paper is the following,
\begin{eqnarray}
 X_1 = \frac12 \rho_{\rm c} S_1 \ , \quad 
 X_2 = \frac12 \rho_{\rm c} S_2 \ , \quad 
X_3 = \rho_3 S_3 \ ,
\label{repcy}
\end{eqnarray}
where the entries of the matrices $S_i$ are given as 
\begin{eqnarray}
&&(S_1)_{mm'}=\delta_{m+1,m'}+\delta_{m-1,m'} \ ,\quad \\
&&(S_2)_{mm'}=i\delta_{m+1,m'}-i\delta_{m-1,m'} \ ,\quad \\
&&(S_3)_{mm'}=(m-1/2)\delta_{m,m'} \ .
\end{eqnarray}
Here the matrix indices $m$ and $m'$ run from $-\infty$ to $+\infty$,
and so the representation is infinite dimensional. 
Note that the dimensionful parameter $\rho_{\rm c}$ does not 
appear in the algebra, thus a family of the representations is
parameterized by a continuous $\rho_{\rm c}$. The physical meaning of
this parameter  is found in the following matrix equation
\begin{eqnarray}
 X_1^2 + X_2^2 = \rho_{\rm c}^2 \identity,
\label{radius}
\end{eqnarray}
which shows that the D0-branes are expected to be on a cylindrical
locus whose radius is $\rho_{\rm c}$, as in the same sense as the
spherical locus relation (\ref{locus}) in the fuzzy $S^2$ case.
The cylinder would be infinitely
long since the range of the eigenvalues of $X_3$ covers the whole region
from $-\infty$ to $+\infty$. In this section, we would like to see if
these expected shape of the cylinder is actually generated in the RR
charge distributions.

\subsection{Isometry and continuum limit}
\label{5.2}

To make sense of the D-brane charge distribution generated from 
matrix configurations of D0-branes, in the previous sections 
we have learned that a simple resolution is to take the continuum
limit. This 
continuum limit is equivalent to the large $N$ limit in the previous
sections, but here for the fuzzy cylinder the representation is already
infinite dimensional. But this doesn't necessarily mean that we need not
to take any limit further. The essence of the large $N$ limit in the
previous sections was that it was the continuum limit in the sense of
the membrane regularization --- 
in that limit all the commutators in the relevant algebra vanish. For
the fuzzy $S^2$ in section \ref{sec3}, the right hand side of the
commutator (\ref{su2x}) has a factor  $1/N$ and thus this continuum
limit is equivalent to the large $N$ limit with fixed $R$. 
In the present case of the fuzzy cylinder, thus we need to
take the limit $\rho_3\rightarrow 0$. For our later convenience, we
write $\rho_3 = l_3/M$ and take the large $M$ limit while $l_3$ is left
fixed. 

The matrix configuration (\ref{repcy}) has a discrete translation
symmetry along the $x_3$ direction by a shift unit $\rho_3$. 
This can be seen as follows. Consider a shift matrix $U$ whose entry is
defined as
\begin{eqnarray}
 U_{mm'} = \delta_{m+1,m'} \ . 
\end{eqnarray}
This $U$ is unitary.
Then the translation along $x_3$ direction by $\rho_3$ can be
represented by this shift operation as
\begin{eqnarray}
 X_i + \delta_{i,3}\identity\rho_3 = U^{-1} X_i U \ .
\end{eqnarray}
This means that any charge density is invariant under this discrete
translation.  
In particular, if we take the large $M$ continuum limit,
the unit of the translation goes to zero and so we may allow any amount
of translation, which means that the charge densities have an isometry
along $x_3$. 

\subsection{D-brane charges}
\label{5.3}

Let us evaluate D-brane charge of the fuzzy cylinder configuration
(\ref{repcy}), in the continuum limit. The distribution before the limit
is taken will be studied in the next subsection. 

\subsubsection{D0-brane charge}
\label{5.3.1}

We evaluate the D0-brane charge distribution (\ref{dens0}).
First, consider the D0-brane density which is integrated over the
$x_1$ and $x_2$ directions, by putting $k_1 = k_2 =0$. 
\begin{eqnarray}
 \widetilde{J}_0(k_1=k_2=0, k_3) = 
{\cal T}_{\rm D0} \sum_m \left[e^{ik_3 (m+1/2)\rho_3}\right] \ .
\end{eqnarray}
After the Fourier transformation, we obtain 
\begin{eqnarray}
 \int \! dx_1 dx_2 J_0(\vec{x}) = {\cal T}_{\rm D0} \sum_m \delta(x_3-
(m+1/2)\rho_3) \ .
\label{afwe}
\end{eqnarray}
If we take the large $M$ limit, we have a continuous uniform
distribution of D0-branes, 
\begin{eqnarray}
\lim_{M\rightarrow \infty}
\frac{1}{M} \int \! dx_1 dx_2 J_0(\vec{x}) 
= \frac{{\cal T}_{\rm D0}}{l_3} \ .
\end{eqnarray}
The density per a unit length along $x_3$ is $M/l_3$. 

To see the full spatial dependence of the 
charge distribution in the large $M$ limit,
we may put $k_3=0$ since we already know that the distribution is
translationally symmetric along $x_3$ in this limit. 
To compute the density 
\begin{eqnarray}
 \widetilde{J}_0(k_1, k_2, k_3=0) = {\cal T}_{\rm D0} {\rm tr} 
\left[e^{ik_1X_1 + i k_2 X_2}\right] \ ,
\end{eqnarray}
we note that this expression is invariant under the rotation around the
axis $x_3$. So we may put $k_2=0$ and recover its $\vec{k}$ dependence
by just replacing $k_1$ by $k_\rho\equiv \sqrt{k_1^2 + k_2^2}$. Then 
\begin{eqnarray}
 \widetilde{J}_0(k_1, k_2, k_3=0) &=& 
{\cal T}_{\rm D0} {\rm tr}
\left[
e^{ik_\rho (\rho_{\rm c}/2)S_1}
\right]
\nn \\
&=&
{\cal T}_{\rm D0} ({\rm tr}\identity)
\sum_{n {\rm \; : \;  even} }
\frac{1}{n!} \left(\frac{ik_\rho\rho_{\rm c}}{2}\right)^n
{}_n C_{n/2} 
\nn \\
&=&
{\cal T}_{\rm D0} ({\rm tr}\identity)\;
B_0(k_\rho\rho_{\rm c}) \ ,
\label{bessel}
\end{eqnarray}
where $B_0(\tau)$ is a Bessel function. Using an integral expression of
the Bessel function, 
\begin{eqnarray}
 B_0(\tau) = \frac{2}{\pi}\int_0^1\! \frac{\cos(\tau s)}{\sqrt{1-s^2}}ds
= \frac{1}{\pi}\int_0^\pi\!d\theta \; e^{i\tau\cos\theta} \ , 
\end{eqnarray}
and the Fourier transform of the delta function of 
$\rho = \sqrt{x_1^2 + x_2^2}$,
\begin{eqnarray}
 \int \! dx_1 dx_2 \; e^{ik_1 x_1 + i k_2 x_2} \delta(\rho-a)
= 2a \int_0^\pi\!d\theta \; e^{ik_\rho a\cos\theta} \ , 
\end{eqnarray}
we obtain the inverse Fourier transform of the density as 
\begin{eqnarray}
 \int dx_3 J_0(\vec{x}) = 
{\cal T}_{\rm D0}({\rm tr}\identity)\;
\frac{1}{2\pi\rho_{\rm c}} 
\delta(\rho-\rho_{\rm c}) \ .
\label{weobthe}
\end{eqnarray}
Using the fact that the integration over $x_3$ is reduced to a
multiplication of the factor ${\rm tr}\identity \rho_3$ in the large $M$
limit, we may write the above expression as a local one, 
\begin{eqnarray}
\lim_{M\rightarrow \infty}
\frac{1}{M}J_0(\vec{x}) = 
{\cal T}_{\rm D0}
\frac{1}{2\pi\rho_{\rm c}l_3} 
\delta(\rho-\rho_{\rm c}) \ .
\label{resultd0}
\end{eqnarray}
This expression shows that the D0-branes are distributed on a cylinder 
with radius $\rho_{\rm c}$. The cylinder is extended along $x_3$ in the
three dimensional space.

\subsubsection{D2-brane charge}
\label{5.3.2}

Next, let us compute the D2-brane charge.
The orientation of the generated D2-brane is along the $x_3$ axis, 
because the following component of the D2-brane charge expression
\begin{eqnarray}
 {\rm tr}\left[[X_1,X_2] e^{ikX}\right]
\end{eqnarray}
vanishes due to the algebra of the fuzzy cylinder, (\ref{cyal}). 

Let us compute the 023 component of the D2-brane charge 
\begin{eqnarray}
 \widetilde{J}_{023}(\vec{k})
= \frac{{\cal T}_{\rm D0}}{2\pi\alpha'}
{\rm tr}\left[-i[X_2,X_3]e^{ik_iX_i}\right]
= \frac{{\cal T}_{\rm D0}\rho_3}{2\pi\alpha'}
{\rm tr}\left[X_1e^{ik_iX_i}\right].
\end{eqnarray}
Again we put $k_3=0$ to evaluate the charge density.
Due to the fact that the expression is covariant
with respect to the rotation around the $x_3$ axis, the evaluation of
the trace should give a function of the form $k_1 f(k_\rho)$. Hence we
may put $k_2=0$ and later will recover the $k_i$ dependence. For
$k_2=0$, 
\begin{eqnarray}
 {\rm tr} \left[
X_1 e^{ik_1 X_1}
\right] 
&=&
\frac{\rho_{\rm c}}{2}{\rm tr} 
\sum_{n=0}^{\infty}\frac{1}{n!} 
\left(\frac{ik_1 \rho_{\rm c}}{2}\right)^n(S_1)^{n+1}
\nn\\
&=&
\frac{\rho_{\rm c}}{2} ({\rm tr}\identity)
\sum_{n\; {\rm : \; odd}>0}^{\infty}\frac{1}{n!} 
\left(\frac{ik_1 \rho_{\rm c}}{2}\right)^n{}_{n+1}C_{(n+1)/2}
\nn\\
&=&
({\rm tr}\identity)
\frac{1}{ik_1}\sum_{j=1}^{\infty}\frac{2j}{(j!)^2}
\left(-\frac{k_1^2\rho_{\rm c}^2}{4}\right)^j
\nn\\
&=&
({\rm tr}\identity)
\frac{1}{i}\frac{\p}{\p k_1} B_0(k_1\rho_{\rm c}) \ .
\end{eqnarray}
So, recovering the covariance, we obtain
\begin{eqnarray}
 \widetilde{J}_{023}(k_1, k_2, k_3=0)
=\frac{{\cal T}_{\rm D0}\rho_3}{2\pi i \alpha'}  
\frac{\p}{\p k_1} J_0(k_\rho\rho_{\rm c}).
\label{recover}
\end{eqnarray}
The Fourier transform is given by
\begin{eqnarray}
 \int \! dx_3\;J_{023}
=\frac{{\cal T}_{\rm D0}\rho_3}{2\pi\alpha'}  ({\rm tr}\identity)
 \frac{x_1}{2\pi\rho_{\rm c}} \delta (r-\rho_{\rm c}).
\end{eqnarray}
This shows that the local D2-brane charge is generated on the cylinder.
In the large $M$ limit, the D2-brane charge distribution obtains the
isometry along $x_3$, thus we are led to the following local expression
\begin{eqnarray}
\lim_{M\rightarrow \infty}J_{023}
=\frac{{\cal T}_{\rm D0}}{2\pi\alpha'}
 \frac{x_1}{2\pi\rho_{\rm c}} \delta (r-\rho_{\rm c}) \ .
\label{localcy}
\end{eqnarray}
This coincides with what we have expected. 
The single D2-brane forms a complete continuous cylinder with
the radius $\rho_{\rm c}$ oriented along $x_3$.
In fact, the total D2-brane charge can be computed as 
\begin{eqnarray}
\int dx_3 \int \! 2\pi\rho d\rho 
\left(
J_{023}\hat{x}_1 - J_{013}\hat{x}_2
\right)
=\frac{{\cal T}_{\rm D0}}{4\pi^2\alpha'}
\cdot  ({\rm tr}\identity)
 \rho_3 \cdot 2\pi\rho_{\rm c}.
\end{eqnarray}
Since $({\rm tr}\identity)\rho_3$ is the length of the cylinder
(which is infinitely long), we can rewrite this as 
${\cal T}_{\rm D2}$ times the total area of the cylinder.

\subsection{From fuzzy $S^2$ to fuzzy cylinder}
\label{5.4}

In the previous subsection we made full use of the isometry in the
continuum limit, but our concern is distribution of the charges 
before the continuum limit is taken. For this purpose we can use the
fact that  
the fuzzy cylinder can be obtained as a special large $N$ limit of an
ellipsoidal deformation of the fuzzy $S^2$. 
In this subsection we consider this deformation to obtain the 
charge distribution of the fuzzy cylinder for $M=1$, 
and see how the results obtained 
in this section are consistent with the results found in section
\ref{sec3}.  

We generalize the fuzzy sphere (\ref{solD0s2}) 
to a fuzzy membrane of an ellipsoidal shape,  as
\begin{eqnarray}
 X_i = \frac{2\alpha_iR}{N} L_i
\end{eqnarray}
in which we introduced asymmetry factors $\alpha_i$ ($i=1,2,3$) 
for each of the three spatial directions.
With this deformation the algebra (\ref{su2x}) is modified accordingly
to
\begin{eqnarray}
 [X_1, X_2] = \frac{2R}{N}
\frac{\alpha_1 \alpha_2}{ \alpha_3} i X_3, \;
 [X_2, X_3] = \frac{2R}{N}
\frac{\alpha_2 \alpha_3}{ \alpha_1} i X_1, \;
 [X_3, X_1] = \frac{2R}{N}
\frac{\alpha_3 \alpha_1}{ \alpha_2} i X_2 \ . 
\label{defo}
\end{eqnarray}
To obtain the algebra of the fuzzy cylinder (\ref{cyal}) from that of
the deformed fuzzy sphere (\ref{defo}), we take the large $N$ limit with 
fixed $\alpha_1$ and $\alpha_2$ as 
\begin{eqnarray}
\alpha_3 = N/2 \rightarrow \infty \ , \quad 
R = \rho_3 \ .
\label{limitN-1}
\end{eqnarray}
In this limit, the spherical locus relation of the fuzzy sphere
(\ref{locus}) is reduced to that
of the fuzzy cylinder (\ref{radius}), if we take
\begin{eqnarray}
\alpha_1 = \alpha_2=\rho_{\rm c}/\rho_3 \ .
\label{limitN-2}
\end{eqnarray}
Note that this large $N$ limit does not make the algebra trivial. To
have a continuum limit we should further take the large $M$ limit with 
$l_3\equiv \rho_3 M$ fixed.

The explicit expression of the $N$ dimensional representation
of $SU(2)$ is (\ref{l3}) and 
\begin{eqnarray}
(L_+)_{mm'} \equiv (L_1+iL_2)_{mm'}
&=& \sqrt{(m'-1)(N-m'+1)}\delta_{m+1,m'} \ ,
\\
(L_-)_{mm'} 
 \equiv (L_1-iL_2)_{mm'}
&=& \sqrt{(N-m')m'}\delta_{m-1,m'} \ .
\end{eqnarray}
In the large $N$ limit (\ref{limitN-1})(\ref{limitN-2}), 
we concentrate on a part of the entries with $m,m' \sim N/2$ (center
parts of the matrices), 
then we can reproduce the representation (\ref{repcy}) of the fuzzy
cylinder algebra. 

The D0-brane charge density before the continuum limit is taken 
can be evaluated precisely in the same way as in section \ref{sec3}. 
The only difference is that the three spatial directions have
different scale factors $\alpha_i$. The D0-brane charge density 
is given as 
\begin{eqnarray}
 \widetilde{J}_0(k_i) 
= {\cal T}_{\rm D0}{\rm tr}\left[e^{ik'_i (2R/N)L_i}\right]
\end{eqnarray}
in which $k'_i \equiv \alpha_i k_i$. 
Then from the rotational invariance in this scaled momentum space, 
we obtain  
\begin{eqnarray}
 \widetilde{J}_0(k_i) = {\cal T}_{\rm D0}\!\!\!\!
\sum_{m=-(N-1)/2}^{(N-1)/2} 
\exp\left[
ik'm\frac{2R}{N}
\right].
\end{eqnarray} 
We have defined 
$k' \equiv |\vec{k'}| =
\sqrt{\alpha_1^2k_1^2 + \alpha_2^2k_2^2 +\alpha_3^2k_3^2}$.
The inverse Fourier transform of this for even $N$ for example is
obtained as 
\begin{eqnarray}
J_0(\vec{x})= 
\frac{1}{\alpha_1\alpha_2\alpha_3}{\cal T}_{\rm D0}
\sum_{n=1}^{N/2}D(r',R^{(n)})
\label{cde}
\end{eqnarray}
where $R^{(n)}$ is defined in (\ref{defrn}) and 
\begin{eqnarray}
r'&\equiv&
\sqrt{(x_1/\alpha_1)^2\!+\!(x_2/\alpha_2)^2\!+\!(x_3/\alpha_3)^2 } 
=\sqrt{
\frac{(\rho_3)^2}{(\rho_{\rm c})^2} \left((x_1)^2+(x_2)^2\right)
+\frac{4}{N^2}(x_3)^2
}
\ . \;\;\;
\label{rp}
\end{eqnarray}
This shows that the shells are deformed ellipsoidally with the scale
factors $\alpha_i$, for finite $N$. 

Let us take the large $N$ limit to obtain the fuzzy cylinder. 
We consider first the case $\rho^2=(x_1)^2 + (x_2)^2\neq 0$. 
The large $N$ limit makes the $x_3$
dependence in $r'$ vanish, then $r' = \rho\rho_3/\rho_{\rm c}$.
In the large $N$ limit the summation over the shells can be replaced by
an integration, and a computation gives
\begin{eqnarray}
 J_0^{(\rho\neq 0)}(x) 
= \frac{{\cal T}_{\rm D0}}{2\pi} \frac{\rho_3}{\rho_{\rm c}}
\left[
\delta(r'-\rho_3) - \lim_{y\rightarrow 0}\frac{\rho_3}{y}\delta(r'-y)
\right] \ .
\end{eqnarray}
In two dimensions the second term is nontrivial (while 
in three dimensions in
section \ref{sec3} the corresponding term is an artifact of the
spherical coordinates) and actually rewritten as 
\begin{eqnarray}
 J_0^{(\rho\neq 0)}(x) 
= \frac{{\cal T}_{\rm D0}}{2\pi\rho_3\rho_{\rm c}} 
\delta(\rho-\rho_{\rm c})  
- \frac{{\cal T}_{\rm D0}}{\rho_3} \delta(x_1)
\delta(x_2) \ .
\label{j01}
\end{eqnarray}
The argument of the second term should be understood as a
limit of approaching to the origin. 
Interestingly, the integration of $J^{(\rho\neq 0)}_0(x)$ over
$x_1$ and $x_2$ vanishes, 
\begin{eqnarray}
 \int \! dx_1 dx_2 \; J_0^{(\rho\neq 0)}(x) = 0.
\end{eqnarray}
The first term of (\ref{j01}) gives a continuous cylindrical
distribution of the D0-branes while the second term is anti-D0-branes
smeared on the $x_3$ axis. 

On the other hand, when $x_1=x_2=0$, the situation is very different. 
Since $r'=2|x_3|/N$, the summation over the shells cannot be replaced by
the integration. 
To look closely at the behaviour around the polar origin $x_1=x_2=0$, 
we consider the case of $x_1 \sim x_2 \sim {\cal O}(1/N)$, and 
introduce a new ${\cal O}(1)$ variable 
$\tilde{\rho}^2\equiv (N^2/4)(\rho_3/\rho_{\rm c})^2[(x_1)^2+(x_2)^2]$. 
Then, $r'$ becomes of order ${\cal O}(1/N)$, 
\begin{eqnarray}
 r' = (2/N)\tilde{r}' \ , \quad \tilde{r}'\equiv
\sqrt{\tilde{\rho}^2+ (x_3)^2} \ .
\end{eqnarray}
After a straightforward calculation one finds
\begin{eqnarray}
J_0^{(\rho\sim 0)}(x)\!
=\!
\frac{{\cal T}_{\rm D0}}{2\pi(\rho_{\rm c})^2}
\!\!\sum_n \frac{-N^2}{(2n\!-\!1)^2}
\left[
\delta\!
\left(\!\tilde{r}'\!-\!\left(\!n\!-\!\frac12\!\right)\!\rho_3\!\right)
+ \rho_3\!\left(\!n\!-\!\frac12\!\right)\!
\delta'\!
\left(\!\tilde{r}'\!-\!\left(\!n\!-\!\frac12\!\right)\!\rho_3\!\!\right)
\right]\!.
\nn
\end{eqnarray}
It is obvious that this term gives a diverging contribution at 
$x_3\sim (n-1/2)\rho_3$ and $x_1\sim x_2\sim 0$, so we compute an
integration of this over a small region in the $x_1$-$x_2$ plane around 
the origin $x_1=x_2=0$. Using the calculation (\ref{strange}), we
obtain  
\begin{eqnarray}
 \int_{x_1, x_2 \sim 0}\! dx_1 dx_2 \; J^{(\rho\sim 0)}_0(x) = 
\sum_n {\cal T}_{\rm D0}
\delta\left(x_3-\left(n-\frac12\right)\rho_3\right)
 \ . 
\end{eqnarray}
Therefore we may conclude that the local expression of this 
is precisely a delta function in the $x_1$-$x_2$ space, as 
\begin{eqnarray}
  J_0^{(\rho\sim 0)}(x_1\!\sim\! 0, x_2\!\sim\!0, x_3) 
={\cal T}_{\rm D0} \sum_n \delta(x_1) \delta(x_2)\;
 \delta\left(x_3-\left(n-\frac12\right)\rho_3\right) \ ,
\end{eqnarray}
in the large $N$ limit.
This result $J^{(\rho\sim 0)}_0(x)$ shows the presence of a 
sequence of un-smeared
D0-branes on the $x_3$ axis.

Adding the two contributions we obtain the full D0-brane distribution, 
$J_0 = J_0^{(\rho\neq 0)} + J_0^{(\rho\sim 0)}$. This full expression is
consistent with the results in the previous subsection, (\ref{afwe}) and
(\ref{weobthe}),
when integrated over a part of the spatial directions.
Of course, in the continuum limit $M\rightarrow \infty$, 
it is easy to show that the result coincides with (\ref{resultd0}) : 
\begin{eqnarray}
 \lim_{N,M\rightarrow \infty} \frac{1}{M} 
\frac{1}{\alpha_1\alpha_2\alpha_3}{\cal T}_{\rm D0}
\sum_{n=1}^{N/2}D(r',R^{(n)})
=
{\cal T}_{\rm D0}
\frac{1}{2\pi\rho_{\rm c}l_3} 
\delta(\rho-\rho_{\rm c}) \ .
\end{eqnarray}
In the continuum limit, the second term in $J^{(\rho\neq 0)}$ is
cancelled completely by 
$J^{(\rho\sim 0)}$, and only the first term in $J^{(\rho\neq 0)}$ is
left, which is the continuous cylinder. 

We conclude that also for the fuzzy sphere, although the representation
is infinite dimensional, the D-brane 
charge distribution is very strange and does not agree with the dual
D2-brane picture. Without taking the continuum limit, we find smeared
anti-D0-brane and scattered D0-branes on the $x_3$ axis.

The local expression of the D2-brane charge (\ref{localcy}) is
reproduced in the same manner, in the large $N, M$ limit. 
One finds that the fuzzy cylinder without taking the continuum limit has
the same D2-brane charge distribution as that of the continuum limit,
interestingly. 

\subsection{Fuzzy supertube}
\label{fst}

So far in section \ref{sec3}, \ref{sec4} and \ref{sec5},
we have concentrated on the determination of the
RR charge distribution from a given fuzzy configuration of D0-branes. 
But in actual situations, one has to solve equations of motion of
the effective D0-brane action to obtain those fuzzy configurations as
classical solutions. 
Various fuzzy solutions are known to emerge.
For example, the dimension $N$ fuzzy $S^2$ solution 
appears in the background constant RR four-form field strength
\cite{myers}, and this static fuzzy solution is expected to have 
a ``dual'' description --- a spherical D2-brane on which there are
$N$ units of magnetic field to represent the bound $N$ D0-branes. 
In this subsection, 
we present a more intuitive way to realize the fuzzy cylinder which
we considered in the previous subsections, as a classical solution. 
It turns out that the resultant
configuration is a fuzzy supertube \cite{supertube}. 

To obtain extended fuzzy objects in the bulk spacetime as static
classical solutions, one has to
introduce some mechanism which keeps the extended state stable. 
In the case of the above fuzzy $S^2$, 
it is due to the external background field strength.   
One of the other ways to obtain the fuzzy spheres is via the Nahm
construction of the monopole, in which the fuzzy $S^2$ appears as a part
of a fuzzy funnel which is a fuzzy expansion of D-strings and
represents the D-strings ending on D3-branes \cite{funnel}. In this case
the stability 
is achieved by introducing nontrivial boundary conditions 
at the end of the worldvolume of the D-strings.

A naive way to stabilize extended structures without turning on any
background is to introduce a
mechanical angular momentum. This makes it possible to have fuzzy
structure with axial symmetry, as solutions of the equations of motion
for  D0-branes:
\begin{eqnarray}
 -\ddot{X_i} + \frac{1}{(2\pi\alpha')^2}[X_j,[X_i,X_j]]=0,
\qquad
 [X_i, \dot{X_i}]=0.
\end{eqnarray}
The second equation is the Gauss low, the dot denotes the time
derivative, and we took the gauge $A_0=0$. 
A solution with the angular momentum in the $x_1$-$x_2$ plane is 
\begin{eqnarray}
 X_1 &=& X_1^{(0)}\cos \omega t  + X_2^{(0)}\sin \omega t \ , \nn\\
 X_2 &=& -X_1^{(0)}\sin \omega t  + X_2^{(0)}\cos \omega t \ ,
\label{rotate}\\
 X_3 &=& X_3^{(0)} \ ,
\nn
\end{eqnarray}
where $X_i^{(0)}$ are the matrices satisfying the fuzzy cylinder
algebra (\ref{cyal}). In the following, we shall use the explicit
representation of the fuzzy cylinder, (\ref{repcy}). Immediately we
find that to satisfy the equations of motion we have to take 
\begin{eqnarray}
 \omega = \frac{\rho_3}{2\pi\alpha'} \ .
\label{omega}
\end{eqnarray}
It is interesting that the radius of the cylinder, $\rho_{\rm c}$, is
not 
related to the angular velocity $\omega$. This reminds us of a supertube 
\cite{supertube} which has a degree of freedom to deform arbitrarily the
shape of its cross section. The arbitrariness of $\rho_{\rm c}$ is
related to this degrees of freedom of the supertube.
As seen below, actually our rotating cylinder has a supersymmetry. 
The solution (\ref{rotate}) is also a solution of the 
following BPS equations of the D0-brane system \cite{BakLee}, 
\begin{eqnarray}
&& [X_1, X_2]=0 \ , \quad D_0 X_3=0 \ , \nn\\
&& D_0X_1 \pm i [X_3, X_1]=0 \ , \quad 
 D_0X_2 \pm i [X_3, X_2]=0 \ .
\end{eqnarray} 
The authors of \cite{BakLee} took a different gauge $X_3=A_0$
to solve these BPS equations. Our solution is just in a gauge different
from theirs, and has a clear mechanical origin. 

\subsubsection{RR densities}
\label{5.5.2}

Let us evaluate the charges generated by the solution (\ref{rotate}),
to see if this
configuration of the rotating D0-branes is actually a supertube
consisting of a cylindrical D2-brane. 
We consider only the continuum limit of the charge distribution, since
without this we would get strange charge distribution as found in 
subsection \ref{5.4}.

First, we evaluate densities of the RR charges. In evaluation, we
can use the results of the previous subsections. Since the charge
density 
formulas of the D0-branes (\ref{dens0}) and
D2-branes (\ref{dens2}) do not include any time derivative, the results
for the rotating solution (\ref{rotate}) can be obtained by simply
replacing $k_1$ by $(k_1 \cos\omega t - k_2 \sin \omega t)$ and 
$k_2$ by $(k_1 \sin\omega t + k_2 \cos \omega t)$ in (\ref{bessel}) and  
(\ref{recover}). However, this
replacement does not change the magnitude of the momentum 
$k_\rho=\sqrt{k_1^2 + k_2^2}$. Hence we
conclude that the D2-brane and D0-brane charge densities in the rotating
case are the same as those of the previous subsections. 
This shows that in the large $N,M$ limit the RR charges are generated
at the precise location needed for the identification with the D2-brane
supertube with the radius $\rho_{\rm c}$.

The D0-brane motion is expected to be along the tube on
which those D0-branes are smeared, so we can never see the motion of the
D0-brane by looking just at the time component of the D0-brane density
$J_0(x)$. However, even in this kind of situation we can extract the
information of the velocity of the D0-branes by looking at the other
components of the D0-brane density,  
\begin{eqnarray}
 \widetilde{J}_i(k) \equiv {\cal T}_{\rm D0} {\rm Str} \left[
\dot{X}_ie^{ik_iX_i}
\right] \ .
\label{v}
\end{eqnarray}
The reason is obvious. The coupling of a single D0-brane to its relevant
RR gauge field is 
\begin{eqnarray}
S_{\rm RR}/{\cal T}_{\rm D0}=  \int \! C^\mu dX_\mu
=  \int \! C^\mu(X(\tau)) \p_\tau X_\mu(\tau) d\tau \ . 
\end{eqnarray}
If we take the static gauge $X_0=\tau$ and turn on a velocity for the
D0-brane as $\p_\tau X_1=v_1$, then the above coupling is written as 
\begin{eqnarray}
S_{\rm RR}/{\cal T}_{\rm D0}
= \int \! C_0(X) d\tau + \int C_1(X) v_1 d\tau.
\end{eqnarray}
So, once all the components of the D0-brane density 
$J_\mu(x)\equiv \delta S_{\rm RR} / \delta C_\mu$ are known,   
one can compute the velocity of the D0-branes by the formula
$v_i = J_i(x)/J_0(x)$. 

The evaluation of (\ref{v}) is almost the same as the evaluation of the
D2-brane density, and the final result is (we put $k_3=0$ as before)
\begin{eqnarray}
\int \! dx_3 \;J_1(x) 
&=& \omega x_2 {\cal T}_{\rm D0} ({\rm tr}\identity) 
\frac{1}{2\pi\rho_{\rm c}} \delta(\rho-\rho_{\rm c}) \ , \\
\int \! dx_3 \;J_2(x) 
&=& -\omega x_1 {\cal T}_{\rm D0} ({\rm tr}\identity) 
\frac{1}{2\pi\rho_{\rm c}} \delta(\rho-\rho_{\rm c}) \ , \\
J_3(x) &=& 0 \ .
\end{eqnarray}
So the velocity is given as
\begin{eqnarray}
 \vec{v} 
= \left(\frac{J_1}{J_0}, \frac{J_2}{J_0}, \frac{J_3}{J_0}\right)
= (\omega x_2, -\omega x_1, 0) \ .
\end{eqnarray}
This shows that the D0-branes forming the cylinder are rotating on the
surface with the angular velocity $\omega$, as expected. This result
might hold even for finite $M$.

\subsubsection{NSNS density and supersymmetry}
\label{5.5.3}

Supertubes have F-string charge as well as the D-brane
charges, and from the dual picture 
we expect that this F-string charge is generated on the
cylinder along the direction of the $x_3$ axis. The F-string density
formula (\ref{Fch}) is evaluated in our case as 
\begin{eqnarray}
 \widetilde{I}_{03}(\vec{k}) 
&=& \frac{{\cal T}_{\rm D0}}{2\pi\alpha'}
{\rm Str} \left[
i[X_3,X_1]\dot{X}_1 e^{ik_iX_i}
+i[X_3,X_2]\dot{X}_2 e^{ik_iX_i}
\right]
\nn\\
&=& \frac{{\cal T}_{\rm D0}}{2\pi\alpha'}\omega\rho_3
{\rm Str} \left[
(X_1 X_1 + X_2 X_2) e^{ik_iX_i}
\right].
\end{eqnarray}
As before, we may put $k_3=0$, then due to the commutation relation
$[X_1, X_2]=0$ we may replace Str by the normal trace. 
Using the relation (\ref{radius}), we obtain
\begin{eqnarray}
  \widetilde{I}_{03}(k_1, k_2,k_3=0) 
&=& \frac{\omega\rho_3 \rho_{\rm c}^2 }{2\pi\alpha'} 
\widetilde{J}_0(k_1, k_2, k_3=0) \ .
\end{eqnarray}
Hence after the inverse Fourier transformation, 
\begin{eqnarray}
 \int \! dx_3 \; I_{03}(x) = 
\frac{{\cal T}_{\rm D0}}{2\pi\alpha'} \omega\rho_3\rho_{\rm c}^2
\; ({\rm tr} \identity)
\;\frac{1}{2\pi\rho_{\rm c}} \delta (\rho-\rho_{\rm c}) \ .
\end{eqnarray}
It can be shown that the other components of the F-string charge
vanish due to the fuzzy cylinder algebra (\ref{cyal}), therefore
we conclude that
the fundamental strings oriented in the $x_3$ direction 
are generated uniformly on the
cylinder with 
the radius $\rho_c$. Noting that $\rho_3 ({\rm tr}\identity)$ is the
length of the cylinder in the $x_3$ direction, the above result
indicates a local expression in the large $M$ limit
\begin{eqnarray}
I_{03}(x) = 
\frac{{\cal T}_{\rm D0}}{2\pi\alpha'} \omega\rho_{\rm c}^2
\frac{1}{2\pi\rho_{\rm c}} \delta (\rho-\rho_{\rm c}) \ . 
\label{resultf1}
\end{eqnarray}

Let us see that this resultant amount of the fundamental 
string charge is precisely reproducing the corresponding quantities
calculated in the dual D2-brane description of the supertube. 
The supersymmetry condition of cylindrically symmetric
D2 supertubes is simply written as 
\begin{eqnarray}
 |{\bf D}| |{\bf B}|=1 \ ,
\label{supert}
\end{eqnarray} 
where ${\bf D}$ is an electric induction on the D2-brane while ${\bf B}$
is a magnetic field. As is well known, ${\bf B}$ is related to the
D0-brane charge density on the D2-brane, as
\begin{eqnarray}
 {\cal T}_{\rm D2} |{\bf B}| \delta(\rho-\rho_{\rm c})= J_0(x) \ ,
\end{eqnarray}
while $|{\bf D}|$ is related to the F-string charge as 
\begin{eqnarray} 
 {\cal T}_{\rm D2} |{\bf D}| \delta(\rho-\rho_{\rm c})= I_{03}(x) \ .
\end{eqnarray}
We use our results of the charge density (\ref{resultd0}) 
and (\ref{resultf1}), to derive the dictionary
\begin{eqnarray}
 |{\bf B}|=\frac{1}{\rho_3}
\frac{1}{2\pi\rho_{\rm c}} 
\frac{{\cal T}_{\rm D0}}{{\cal T}_{\rm D2}} \ ,
\quad 
 |{\bf D}|=\frac{\omega\rho_{\rm c}^2}{2\pi\alpha'}
\frac{1}{2\pi\rho_{\rm c}} 
\frac{{\cal T}_{\rm D0}}{{\cal T}_{\rm D2}} \ .
\end{eqnarray}
Substituting (\ref{omega}) coming from the equations of motion into
these equations, 
we can show that these quantities actually satisfy the supersymmetry
condition (\ref{supert}).
This provides a proof that our rotating D0-branes in the large $M$
limit form a single D2-supertube.

%%%%%%%%%%%%%%%%%%%%%%%%%%%%%%%%%%%%%%%%%%%%%%%%%%%%%%%%%%%%%%
%%%%%%%%%%%%%%%%%%%%%%%%%%%%%%%%%%%%%%%%%%%%%%%%%%%%%%%%%%%%%%
%%%%%%%%%%%%%%%%%%%%%%%%%%%%%%%%%%%%%%%%%%%%%%%%%%%%%%%%%%%%%%

\section{Corrections to the density formulas and the duality}
\label{sec6}

\subsection{Discussions for the duality with finite $N$}

In section \ref{sec3}, \ref{sec4} and \ref{sec5}, 
we have seen that the RR charge density of 
{\it the continuum limit} of the fuzzy objects coincide with
the dual picture of the spherical (or cylindrical) D2-brane (or
D4-brane) with smeared D0-branes bound on it. This is basically what has
been expected in the sense of the matrix regularization. 
To see this in more detail, we recall the fact that in the continuum
limit of the matrix regularization we may replace the trace of the
matrices by the two dimensional integration over the spatial worldvolume
of the membrane (parameterized by $\sigma_i$) \cite{shimada},  
\begin{eqnarray}
\frac{ \int d^2\sigma M^{\rm cont}(\sigma_i)}{\int d^2 \sigma} \sim
\frac1N {\rm tr} M \ ,
\end{eqnarray}
where $M^{\rm cont}(\sigma)$ is the membrane configuration corresponding
to the large $N$ limit of the matrices $M$. 
This results in the evaluation of $J_0(x)$ in our case as
\begin{eqnarray}
 \int\!\! d^dk\; e^{-ik_i x_i }
\frac1N {\rm tr} 
\left[e^{ik_iX_i} \right]
\sim
\frac{1}{\int d^2 \sigma}
 \int\!\! d^dk \; e^{-ik_i x_i }\!\int\!\! 
d^2\sigma \; e^{ik_i X_i^{\rm cont}(\sigma)}
\sim \delta (x_i\! -\! X_i^{\rm cont}(\sigma)) \ ,
\nn
\end{eqnarray}
which coincides with our results obtained in section \ref{sec3},
\ref{sec4} and \ref{sec5}. In this sense, we can say that 
what we have computed in this paper is a precise version of this
``$\sim$'' relation by explicit evaluations of the trace in the matrix
configurations, for respective supergravity couplings. This ``$\sim$''
makes sense only in the continuum limit where the algebra becomes
trivial.

However, before taking the continuum limit, 
we encounter strange distribution of the RR charges --- the collection
of many spherical shells. Does this finite $N$ (or finite $M$)
configuration make sense for the democratic duality with finite $N$? 

In a certain case, 
it is possible to make sense of the charge distributions
while the fuzziness is kept.
The example is the fuzzy (noncommutative) plane (\ref{ncplane}).
Though the representation is infinite dimensional, the manifold itself 
is non-compact and thus the fuzziness still remains, as seen in the fact
that the D0-brane density in a unit area of the plane is finite
($=1/2\pi\theta$). This is reflected in the fact that, although the
dimension of the representation is infinite, the algebra itself is still
nontrivial.   
In this sense we haven't took the continuum limit for
this fuzzy plane. There is a dual D2-brane description of
this RR charge distribution. It is a flat bound state 
of a D2-brane and smeared D0-branes, which reproduces the 
charge distributions (\ref{ncp1}) and (\ref{ncp2}). 
The resultant brane configuration has a worldsheet
description as a conformally invariant 
boundary of the worldsheet with mixed boundary conditions. From 
this example, we might naively expect that even for finite $N$
(or more precisely, before taking the continuum limit, i.e. when the
algebra is nontrivial), the D0-brane
picture should agree with the dual D2-brane (D4-brane) picture, giving
the same distribution of the RR charges. However, in the example of the
fuzzy $S^2$, we found in section \ref{sec3} that this is obviously
violated. First, there are shells, and 
second, the shells have fractional D2-brane charges. Why in this case of
the fuzzy spheres do the charge distributions disagree with the dual
D2-brane picture?

We discuss possible resolutions to this interesting puzzle. 
Since there is no known worldsheet description of fractional D-branes
without any orbifolding, we believe that the results presented in this
paper should be corrected for finite $N$, somehow. The finite $N$
corrections 
would come from the commutators of $X_i$'s, thus the possible
corrections to the RR charge density formulas are of this commutator
type. There is a naive reason why we may expect commutator corrections :
T-dual of the commutators are covariant derivatives, and  
we may expect the derivative corrections since the density
formulas have been derived at the low energy. So a priori there 
seems to be no reason why we may discard the commutator corrections in
the charge density formulas. This includes the possibility of changing
the ordering prescription in the density formulas.

In some situations with supersymmetries, one might discard the
derivative 
corrections since they are cancelled with each other (see
\cite{Thor}). On the other hand, we may embed the fuzzy 
$S^2$ into a fuzzy funnel \cite{funnel} and in this way the fuzzy
spheres we have treated can be a supersymmetric configuration. 
Therefore we might argue that the fuzzy sphere may receive no
correction. 
However, precisely speaking, there
is no proof that all the higher derivative corrections vanish by the 
preserved supersymmetries, so in a strict sense we cannot rely on the
supersymmetry.  

One of the other possibilities to resolve the discrepancy
is concerning whether the matrix
configuration is on-shell or not. Since the above belief on the duality
of the pictures is based on a unified description a la boundary of 
string worldsheet, and so we might naively guess that 
the duality of the pictures might be present only for a conformally
invariant boundary conditions. This means that, to show the democratic
duality, the D0-brane configurations (and also the dual D2-brane 
configurations) should be on-shell and satisfy their equations of
motion.  
(If the configuration of $X_i$ is off-shell, the
corresponding boundary state (\ref{bs}) may contain divergence
\cite{gbs}.) 
In our situation, it is known that the static 
fuzzy sphere itself is not a
solution of the equations of motion for the D0-branes in a flat
background, and to make it a solution one 
has to introduce a constant RR 4-form field strength background
\cite{myers}, or a near horizon geometry of an NS5-brane, for
example. At the  
core of the NS5-brane, the spacetime is not flat but curved and in fact
becomes $S^3$. The fuzzy $S^2$ appears as a conjugacy class of an
$SU(2)$ ($\sim S^3$) group manifold. If we assume that 
the density formula (\ref{dens0}) persists in this curved background,
naively 
we have to perform the Fourier transformation on $S^3$ (and not $R^3$
which led us to the collections of shells). So it might be possible to
make the shells disappear due to the different metric in the momentum 
space.\footnote{See \cite{sugawara} for a related discussion.} However,
note that if we embed the fuzzy $S^2$ into the fuzzy funnel as discussed
above, it becomes on-shell in a flat spacetime and this kind of argument
doesn't apply.  

One way to see indirectly the geometry of given matrices 
is to calculate the fluctuation spectrum around the matrix configuration
in D0-brane effective field theory and match it
with a spectrum of a string with worldsheet boundary conditions
corresponding to the expected brane configuration. If there really are
the collection of the shells, an interaction of each shell with a probe
D0-brane would give tachyonic fluctuations. 
The fluctuation spectrum of the fuzzy sphere solution with a separated
single D0-brane, corresponding to the present situation, 
has been computed in \cite{KH-Krasnov} 
and the conclusion was that there is a single spherical D2-brane,  
no such collection of the shells. 
However, one should note that 
the spectrum calculation was performed with the 
lowest order action without any higher derivative terms and for a large
$N$.  So this result of the fluctuation analysis 
can receive nontrivial commutator corrections.

In sum, various arguments above suggest that 
in the charge density formulas there should be $1/N$ corrections which
make the strange shells vanish to give a single spherical
D-brane.\footnote{ Even if this puzzle is resolved,
there remains another puzzle : if one starts with a single D0-brane, it
seems impossible to generate extended objects, while in the dual
D2-brane picture, one can easily consider a spherical D2-D0 bound state
with only a single D0-brane charge in total. To show the finite $N$
duality, one has to solve this another puzzle special for the $N=1$ 
case. Through the embedding of the fuzzy sphere into the fuzzy funnel,
this puzzle is related to the fact that the Nahm data becomes
trivial for a monopole with a single magnetic charge. }
Those might be due to the possible commutator (higher derivative or
ordering) 
corrections.\footnote{See \cite{trek3} for a related discussion on
correction terms.} 

\subsection{Corrections to the formulas}

In the following, we shall derive the first
nontrivial commutator correction to the D0-brane charge density formula
(\ref{dens0}), assuming that the corrected density formula gives only a
single shell of the D0-brane distribution for the fuzzy $S^2$ and the
fuzzy $S^4$ with finite $N$.

Let us consider the multipole expansion of the D0-brane charge density.
When $X_i$'s can be diagonalized simultaneously, the formula
(\ref{dens0}) gives a correct result in which the D0-branes are
distributed with the delta-function support on the points specified by
the eigenvalues of the $X_i$. Hence the corrections should be of the
commutator form, and the first nontrivial correction in expansion in
terms of $k$ should be
\begin{eqnarray}
\frac{\widetilde{J}_0(\vec{k})}{{\cal T}_{\rm D0}}=
\left(
{\rm tr} \left[e^{ik_iX_i}\right]
\!+\! c k^4\; {\rm tr}\left[[X_i,X_j]^2\right] t
\!+\! \widetilde{c} k^2\; {\rm tr}\bigl[[X_i,k_lX_l][X_i,k_mX_m]\bigr] 
\!+\! {\cal O}(k^6)\right) \ ,
\label{corr}
\end{eqnarray}
where $c$ and $\widetilde{c}$ are constants which we will determine from 
the assumption mentioned above.  

First we study the fuzzy $S^2$. 
If the resulting D0-brane distribution is a single sphere with its
radius $a$, then (\ref{deltar}) shows that its momentum representation
is 
\begin{eqnarray}
\frac{ \widetilde{J}_0(\vec{k})}{{\cal T}_{\rm D0} }
&=& 
{\cal N} 
\frac{-2\pi ia}{k}\left(e^{ika}-e^{-ika}\right)
\nn\\
&=& 
4\pi a^2{\cal N}
\left[1-\frac{1}{3!}k^2a^2+\frac{1}{5!}k^4a^4+{\cal O}(k^6)\right] \ ,
\end{eqnarray}
where ${\cal N}$ is a normalization constant. 
The value of $\widetilde{J}(k=0)$ gives the total number of the
D0-branes, so we determine ${\cal N}$ as 
${\cal N} = N/4\pi a^2$, then
\begin{eqnarray}
\frac{ \widetilde{J}_0(\vec{k})}{ {\cal T}_{\rm D0} }
=N 
\left[1-\frac{1}{3!}k^2a^2+\frac{1}{5!}k^4a^4+{\cal O}(k^6)\right] \ .
\label{corr1}
\end{eqnarray}
On the other hand, we substitute the fuzzy $S^2$ configuration 
(\ref{solD0s2}) into the corrected D0-brane density (\ref{corr}), we
obtain (for even $N$)
\begin{eqnarray}
\frac{\widetilde{J}_0(\vec{k})}{N{\cal T}_{\rm D0} }
= 
1 -\frac{N^2\!-\!1}{6N^2}k^2R^2
+ \frac{N^2\!-\!1}{N^4}\left(\frac{3N^2\!-\!7}{360}
-32\left(\!c+\frac{\widetilde{c}}{12}\right)\right)k^4 R^4
+{\cal O}(k^6)
\ .
\label{corr2}
\end{eqnarray}
Note that there is no correction of order $k^2$ since no commutator
correction can be found in a way consistent with the dimensionality.
(A possible dimensionful parameter $\alpha'$ should not appear here
since one can bring $R$ to be very large so that those stringy
corrections are negligible.) First let us see that the coefficient of
${\cal O}(k^2)$ determines the radius of the single shell uniquely. 
Comparing (\ref{corr1}) and (\ref{corr2}), we find 
\begin{eqnarray} 
a = R \sqrt{1-\frac{1}{N^2}} \ .
\end{eqnarray}
This is the relation between the physical radius $a$ of the sphere on
which the D0-branes are distributed and the normalization $R$ of the
fuzzy sphere solution (\ref{solD0s2}). Interestingly, using this
relation, we find that the spherical locus relation (\ref{locus}) is
consistent with its physical interpretation without any finite $N$
correction, 
\begin{eqnarray}
 \sum_{i=1}^3 X_i^2 = a^2 \identity_N \ .
\end{eqnarray}
We compare ${\cal O} (k^4)$ term in (\ref{corr1}) and
(\ref{corr2}). We find that they agree if 
\begin{eqnarray}
 c+ \frac{\widetilde{c}}{12} = \frac{-1}{2880} \ .
\label{2880}
\end{eqnarray}

Next, we perform a similar calculation for the fuzzy $S^4$. Assuming
that the D0-brane charge is  generated on a single shell of a
four-sphere, in the same manner we obtain from the Fourier
transformation of a delta function with a support at a radius in the
five dimensions $r=\widetilde{a}$, 
\begin{eqnarray}
\frac{ \widetilde{J}_0(\vec{k})}{{\cal T}_{\rm D0}}
=N 
\left[1-\frac{1}{10}k^2\widetilde{a}^2
+\frac{1}{280}k^4\widetilde{a}^4+{\cal O}(k^6)\right] \ .
\label{corr3}
\end{eqnarray}
Here we have already fixed the overall normalization so that the total
number of the D0-branes is $N$. On the other hand, the corrected formula 
(\ref{corr}) gives with the substitution of the fuzzy $S^4$
configuration (\ref{fuzzys4sol}) 
\begin{eqnarray}
\frac{\widetilde{J}_0(\vec{k})}{N {\cal T}_{\rm D0} }
= 
1 -\frac{4\!+\!n}{10n}k^2R^2
+ \frac{n\!+\!4}{n^3}\left(\frac{3n^2\!+\!12n\!-\!8}{840}
-16\left(c+\frac{\widetilde{c}}{5}\right)\right)k^4 R^4
+{\cal O}(k^6)
\ . \;
\label{corr4}
\end{eqnarray}
To derive this result, we used the rotational invariance of the
expression and also various formulas given in \cite{Castelino}.
The comparison of ${\cal O}(k^2)$ terms in (\ref{corr3}) and
(\ref{corr4}) identifies $R$ in terms of $\widetilde{a}$ as
\begin{eqnarray}
\widetilde{a} = R \sqrt{1+ \frac{4}{n}} \ . 
\end{eqnarray} 
This again shows that the correction factor in 
the four-spherical locus relation (\ref{4locus}) is due to the 
wrong normalization of the matrices $X_i$, 
\begin{eqnarray}
  \sum_{i=1}^{5}X_i^2 = \widetilde{a}^2 \identity_N \ .
\end{eqnarray}
Together with the relation between $R$ and $\widetilde{a}$, the
comparison of (\ref{corr3}) and (\ref{corr4}) at ${\cal O}(k^4)$
provides 
\begin{eqnarray}
 c+ \frac{\widetilde{c}}{5} = \frac{-1}{1680} \ .
\label{1680}
\end{eqnarray}

Hence the assumption that the D0-branes should be distributed on a
single shell for both the fuzzy $S^2$ and the fuzzy $S^4$ can determine
the numerical coefficients of the corrections as
\begin{eqnarray}
 c = \frac{-1}{5880} \ , \quad 
 \widetilde{c} = \frac{-5}{2352} \ .
\end{eqnarray}
These are the first nontrivial corrections to the D0-brane 
charge density formula. 
Note that these $c$ and $\widetilde{c}$  do not depend on $N$, as they
should be since the charge density formulas themselves 
are expected to be independent of the dimension of the 
matrices $X_i$. In the same manner, corrections to the D2-brane and
D4-brane charge density formulas may be obtained. 

It is interesting to study the meaning of these correction terms in more
detail, and obtain higher order terms such as ${\cal O}(k^6)$. 
Unfortunately, our assumption that the fuzzy $S^2$ configuration
(\ref{solD0s2}) (and the fuzzy $S^4$ configuration (\ref{fuzzys4sol}))
may give a single shell in $J_0$ cannot completely 
determine the higher order structure, since there appears many
correction terms due to the rapid growing of the number of the ways of
contracting the indices when a larger number of
commutators is considered. Therefore to fix all the
coefficients we need some other severe physical principle.
It would be intriguing to find it and further investigate 
the duality between the dual pictures with different worldvolume
dimensions.

%%%%%%%%%%%%%%%%%%%%%%%%%%%%%%%%%%%%%%%%%%%%%%%%%%%%%%%%%%%%%%
%%%%%%%%%%%%%%%%%%%%%%%%%%%%%%%%%%%%%%%%%%%%%%%%%%%%%%%%%%%%%%
%%%%%%%%%%%%%%%%%%%%%%%%%%%%%%%%%%%%%%%%%%%%%%%%%%%%%%%%%%%%%%

\acknowledgments

I am indebted to W.~Taylor for 
collaboration at the early stage of this work. I would like to thank
A.~Jevicki,
M.~Hamanaka, A.~Hashimoto, Y.~Hyakutake, M.~Kato, S.~Nagaoka, T.~Sato,
H.~Shimada and T.~Yoneya for fruitful discussions, and also the members
of the particle theory group at Institute of Physics, the university of
Tokyo. I am grateful to the organizers of the workshop on branes and 
generalized dynamics at Argonne National Laboratory where a part of this
work was done. This work was supported in part by the Grant-in-Aid for
Scientific Research (No.~12440060, 13135205, 15540256 and 15740143) from
the Japan Ministry of Education, Science and Culture.

\appendix

\section{Democratic duality from tachyon condensation}
\label{appA}

To find a manifestly democratic description of various fuzzy 
objects is a different issue beyond the
aim of this paper, but in this appendix let us present an
example. For the confirmed duality in the flat D2-D0 bound
state studied in subsection \ref{ncp}, we may find another interesting
description which is manifestly democratic in a sense. We use
non-BPS D1-branes as a starting point and can make a D2-D0 solution
which exhibits the duality.

On parallel $N$ non-BPS D1-branes, we have a real tachyon and
transverse scalar fields in adjoint representation of $U(N)$. 
In the large $N$ limit one can construct a flat BPS D2-brane by
turning on these fields as \cite{terashima}
\begin{eqnarray}
 T = u \hat{p}_2 \ , \quad \Phi_2 = \hat{x}_2 \ ,
\label{sold2pq}
\end{eqnarray}
where $u$ is a constant parameter which is taken to be infinite to
satisfy the equations of motion of a boundary string field theory (BSFT)
for the D1-branes, and 
$\hat{x}_2$ and $\hat{p}_2$ are matrix representation of the
Heisenberg algebra 
\begin{eqnarray}
 [\hat{x}_2, \hat{p}_2] = i \ . 
\label{qp}
\end{eqnarray}
The first description (i) in section \ref{sec1} is a D2-brane with a
constant magnetic field. 
In order to turn on the gauge field on the D2-brane
solution (\ref{sold2pq}), 
we introduce a fluctuation of the tachyon solution as 
\cite{AST}
\begin{eqnarray}
  T = u \bigl(\hat{p}_2 + A_2(\hat{x}_2, x_1, x_0)\bigr) \ .
\end{eqnarray}
This is basically a covariant derivative.
For our concern, we put $A_2 = f x_1$ to turn on the constant field
strength $F_{12}=f$. The resultant configuration 
\begin{eqnarray}
 T = u \bigl(\hat{p}_2 + fx_1\bigr) \ , \quad \Phi_2 = \hat{x}_2 \ ,
\label{solb}
\end{eqnarray}
is again a solution of the BSFT. This field strength generates D0-branes
bound in the D2-brane and their density is determined by $f$ as
explained in subsection \ref{ncp}. 

We notice here that this (\ref{solb}) has
another interpretation which becomes manifest when we rewrite it as 
\begin{eqnarray}
 T = fu \bigl(x_1 - \Phi_1\bigr) \ , \quad \Phi_2 = \hat{x}_2 \ ,
\label{man}
\end{eqnarray}
where $\Phi_1 = -\hat{p}_2/f$. The first equation in (\ref{man}) means 
a collection of D0-branes since $x_1$ is a kink on the non-BPS
D1-branes,  and the location of those D0-branes are specified by
eigenvalues of $\Phi_1$ which give zeros of the tachyon field. 
So, an interpretation of (\ref{solb}) is a collection of the
D0-branes whose location is given by $x_1=\Phi_1$ and $x_2=\Phi_2$. Due
to (\ref{qp}), we obtain
\begin{eqnarray}
 [\Phi_1, \Phi_2] = \frac{i}{f}
\end{eqnarray}
which is what we studied in subsection \ref{ncp}. The factor $1/f$
appears consistently. This coincides the second description (ii) in
section \ref{sec1}. 

Thus we conclude that the solution (\ref{solb})
on the non-BPS D1-branes has two meanings, one is a flat D2-brane with
the constant field strength (i), and another is a collection of
D0-branes satisfying the Heisenberg algebra (ii). The solution
(\ref{solb}) is a simultaneous   
manifestation of these two equivalent descriptions.  

%%%%%%%%%%%%%%%%%%%%%%%%%%%%%%%%%%%%%%%%%%%%%%%%%%%%%%%%%%%%%%
%%%%%%%%%%%%%%%%%%%%%%%%%%%%%%%%%%%%%%%%%%%%%%%%%%%%%%%%%%%%%%
%%%%%%%%%%%%%%%%%%%%%%%%%%%%%%%%%%%%%%%%%%%%%%%%%%%%%%%%%%%%%%

%%%%%%%%%% References %%%%%%%%%%%%%%%%%%%%%%%%%
\newcommand{\J}[4]{{\sl #1} {\bf #2} (#3) #4}
\newcommand{\andJ}[3]{{\bf #1} (#2) #3}
\newcommand{\AP}{Ann.\ Phys.\ (N.Y.)}
\newcommand{\MPL}{Mod.\ Phys.\ Lett.}
\newcommand{\NP}{Nucl.\ Phys.}
\newcommand{\PL}{Phys.\ Lett.}
\newcommand{\PR}{ Phys.\ Rev.}
\newcommand{\PRL}{Phys.\ Rev.\ Lett.}
\newcommand{\PTP}{Prog.\ Theor.\ Phys.}
\newcommand{\hep}[1]{{\tt hep-th/{#1}}}
%%%%%%%%%%%%%%%%%%%%%%%%%%%%%%%%%%%%%%%%%%%%%%%

\end{document}